\documentclass[useAMS,usenatbib]{mn2e}
\usepackage{graphicx}
\usepackage{latexsym,amsmath,amssymb}
\usepackage{subfigure}

\title{Red giant stellar collisions in the Galactic Centre}

\author[James E. Dale, Melvyn B. Davies, Ross P. Church, Marc Freitag]{James. E. Dale$^{1}$\thanks{E-mail: jim@astro.lu.se (JED)},
Melvyn B. Davies$^{1}$, Ross P. Church$^{1,2}$, Marc Freitag$^{3}$\\
$^{1}$Lund Observatory, Box 43, SE--221 00, Lund, Sweden\\
$^{2}$School of Mathematical Sciences, Monash University, Victoria 3800, Australia\\
$^{3}$Institute of Astronomy, Madingley Road, Cambridge, CB3 0HA, UK}

\begin{document}

\pagerange{\pageref{firstpage}--\pageref{lastpage}} \pubyear{2006}

\maketitle

\label{firstpage}

\def\solmas{{M$_\odot$}}
\def\solm{{M_\odot}}
\def\mnras{MNRAS}
\def\apj{ApJ}
\def\aap{A\&A}
\def\aj{AJ}
\def\apjl{ApJL}
\def\apjs{ApJS}
\def\bain{BAIN}
\def\pasp{PASP}
\def\araa{ARA\&A}
\def\physrep{Phys. Rep.}
\def\aaps{AAPS}
\def\ga{\sim}

\begin{abstract}
We show that collisions with stellar--mass black holes can partially explain the absence of bright giant stars in the Galactic Centre, first noted by \cite{1996ApJ...472..153G}. We show that the missing objects are low--mass giants and AGB stars in the range $1-3$\,M$_{\odot}$. Using detailed stellar evolution calculations, we find that to prevent these objects from evolving to become visible in the depleted K bands, we require that they suffer collisions on the red giant branch, and we calculate the fractional envelope mass losses required. Using a combination of Smoothed Particle Hydrodynamic calculations, restricted three--body analysis and Monte Carlo simulations, we compute the expected collision rates between giants and black holes, and between giants and main--sequence stars in the Galactic Centre. We show that collisions can plausibly explain the missing giants in the $10.5<K<12$ band. However, depleting  the brighter ($K<10.5$) objects out to the required radius would require a large population of black hole impactors which would in turn deplete the $10.5<K<12$ giants in a region much larger than is observed. We conclude that collisions with stellar--mass black holes cannot account for the depletion of the very brightest giants, and we use our results to place limits on the population of stellar--mass black holes in the Galactic Centre.
\end{abstract}

 
\section{Introduction}
In common with many other galaxies, the Milky Way hosts at its centre a supermassive black hole, Sagittarius A*, whose mass is $\approx4\times 10^{6}$\,M$_{\odot}$ \citep[e.g.][]{2003ApJ...596.1015S}. Surrounding Sgr A* is the Galactic Centre cluster, the study of which offers unique insights into many areas of astrophysics, such as the dynamics of star clusters with central massive objects, the formation history of supermassive black holes and the emission of gravitational waves by Extreme Mass--Ratio Inspirals (for a recent review of this topic, see \cite{2007astro.ph..3495A}.\\
\indent The study of the Galactic Centre cluster is hindered by $\sim30$\,magnitudes of intervening visual extinction. Nevertheless, a great deal has been learned about the stellar population in the central few parsecs by observations made in X--rays and in the near infra--red, in which the extinction is only $\sim3$\,magnitudes.\\
\indent The Galactic Centre Cluster is one of the densest stellar systems known. \citep{1996ApJ...472..153G} concluded that the Cluster had a dense core with a radius of $\approx 0.38$pc and a density of $4\times 10^{6}$M$_{\odot}$pc$^{-3}$. Later measurements by \cite{2003ApJ...594..812G} and \cite{2007A&A...469..125S} found that the cluster was more cusp-like and reported still higher number densities, reaching $\approx10^{7}$pc$^{-3}$ at a radius of $0.1$pc. These number densities are well in excess of those of globular clusters \citep[$\lesssim 10^{6}$M$_{\odot}$pc$^{-3}$, ][]{1996AJ....112.1487H} or young clusters such as the Arches \citep[$3\times10^{5}$M$_{\odot}$pc$^{-3}$, ][]{1999ApJ...525..750F}. The cusp is composed largely of old and intermediate--age stars \citep[][]{2003ApJ...594..812G,2007A&A...469..125S}. However, mixed with this population are two unusual groups of young stars. The S--stars \citep[e.g.][]{2005ApJ...620..744G}, orbiting within $\sim0.01$\,pc have attracted considerable attention because they appear to be young B--stars living in a volume in which star formation should be strongly suppressed by the tidal field of Sgr A*. Alternative suggestions for the origin of the S--stars include the inward migration and dissolution of a dense stellar cluster \citep[e.g.][]{2001ApJ...546L..39G, 2003ApJ...593L..77H, 2003ApJ...593..352P, 2005ApJ...628..236G} or the disruption of a massive binary injected from larger radii \citep{2003ApJ...592..935G}, in which case they are young but formed elsewhere, and the tidal stripping of giant stars \citep{2005ApJ...624L..25D}, in which case the S--stars are locally formed but not young.\\
\indent Further out, between radii of $\sim0.04$ and $\sim0.3$\,pc, there appear to be two discs of young stars oriented at large angles both to each other and to the Galactic plane \citep{2003ApJ...594..812G,2006ApJ...643.1011P}. There is strong evidence that these discs contain large numbers of massive stars and have an unusually flat IMF \citep{2006ApJ...643.1011P}.\\
\indent In this paper, we investigate a peculiar property of the background cluster of older stars; within $\sim0.2$pc, the Galactic Centre cluster is deficient in bright giant stars. This was first documented by \cite{1996ApJ...472..153G} who observed the Galactic Centre in the near infrared K--band. They used the presence or absence of CO bandhead absorption to distinguish late--type from early--type stars and divided their sample into three K--bands; $12<K<15$, $10.5<K<12$ and $K<10.5$. While the surface density of late--type stars in the faintest of these bands appears to continue smoothly inwards to very small radii, there are clear holes in the distributions of late--type stars in the other bands, with projected radii of $\sim0.08$ and $\sim0.2$\,pc respectively.\\
\indent Since the Galactic Centre is a densely--populated region, \cite{1996ApJ...472..153G} proposed that the missing giants had been destroyed by collisions with main--sequence stars. This scenario was investigated by \cite{1999ApJ...527..835A} and \cite{1999MNRAS.308..257B} but these authors reached different conclusions. \cite{1999ApJ...527..835A} found that collisions with MS stars could explain the observed depletion of giants, while \cite{1999MNRAS.308..257B} concluded that collisions with main sequence stars, white dwarfs and neutron stars would not do sufficient damage to a giant to prevent it becoming visible in the depleted K--bands. Their different conclusions largely arise from the criteria they used to determine whether or not an impact had done sufficient damage to prevent a given giant appearing in the depleted K--bands in the Galactic Centre. \cite{1999MNRAS.308..257B} studied encounters involving two very late giants. They found that the mass losses achieved by collisions with MS and compact--object impactors were small, no greater than $10\%$ of the envelope. They also considered that removal of `most if not nearly all of the envelope' was required to prevent the giant becoming visible in the depleted K--bands (as we will show later, this is largely correct for giants towards the tip of the giant branch). They therefore concluded that collisions with MS stars were unable to explain the depletion of giants. Conversely, \cite{1999ApJ...527..835A} made the assumption that a collision with a single MS impactor would result in the total destruction of a giant's envelope if the ratio of the impact parameter to the giant radius, $x_{\rm e}$, were less than $0.25$, and that an impact with a binary would destroy the giant envelope if $x_{\rm e}<1$. Essentially taking collisions to be more effective than \cite{1999MNRAS.308..257B}, \cite{1999ApJ...527..835A} found that encounters could explain the depletion of giants within a radius of $\sim2$ arcsec. By performing hydrodynamic modelling of encounters between MS impactors and younger, less extended giants than those studied by \cite{1999MNRAS.308..257B}, we improve on and extend the work of both of these authors. \cite{1998MNRAS.301..745D} investigated the possibility that collisions with binary MS stars or neutron stars might be responsible for removing the giants. They concluded that such encounters would not have a significant effect on the giant population in the Galactic Centre unless an unrealistically large proportion of binaries was assumed.\\
\indent On theoretical grounds, a large population of stellar--mass black holes is expected to exist in the Galactic Centre \citep[e.g.][]{1993ApJ...408..496M, 2000ApJ...545..847M, 2006ApJ...649...91F} \citep[for reviews of stellar relaxation processes around SMBH, see][]{2006JPhCS..54..243A, 2007arXiv0708.0688A}. This population is of great interest, but is by its nature extremely difficult to study. In this paper, we explore the possibility of using the effects of collisions involving these black holes to study this population indirectly.\\
\indent We extend the work of \cite{1999MNRAS.308..257B} by considering collisions between giants and stellar--mass black holes, and also by allowing for multiple collisions of a single giant with black holes, MS stars or a combination of the two types of object. We then consider what can be learned from our results about the population of stellar mass black holes that surely exists in the Galactic Centre.\\
\indent In Section 2, we discuss the properties of the Galactic Centre stellar population and identify the stars that are missing. In Section 3, we derive the expected collision rates of giant stars with main--sequence stars and black holes. We describe our numerical methods in Section 4 and in Section 5 we explain how we have modeled the effects of mass loss on giant stars. In Section 6, we present the results of our simulations of collisions of giants with black holes and main--sequence stars. We discuss our results in Section 7 and draw our conclusions in Section 8.\\
\section{Missing stars in the Galactic Centre}
\begin{figure}
\includegraphics[width=0.45\textwidth]{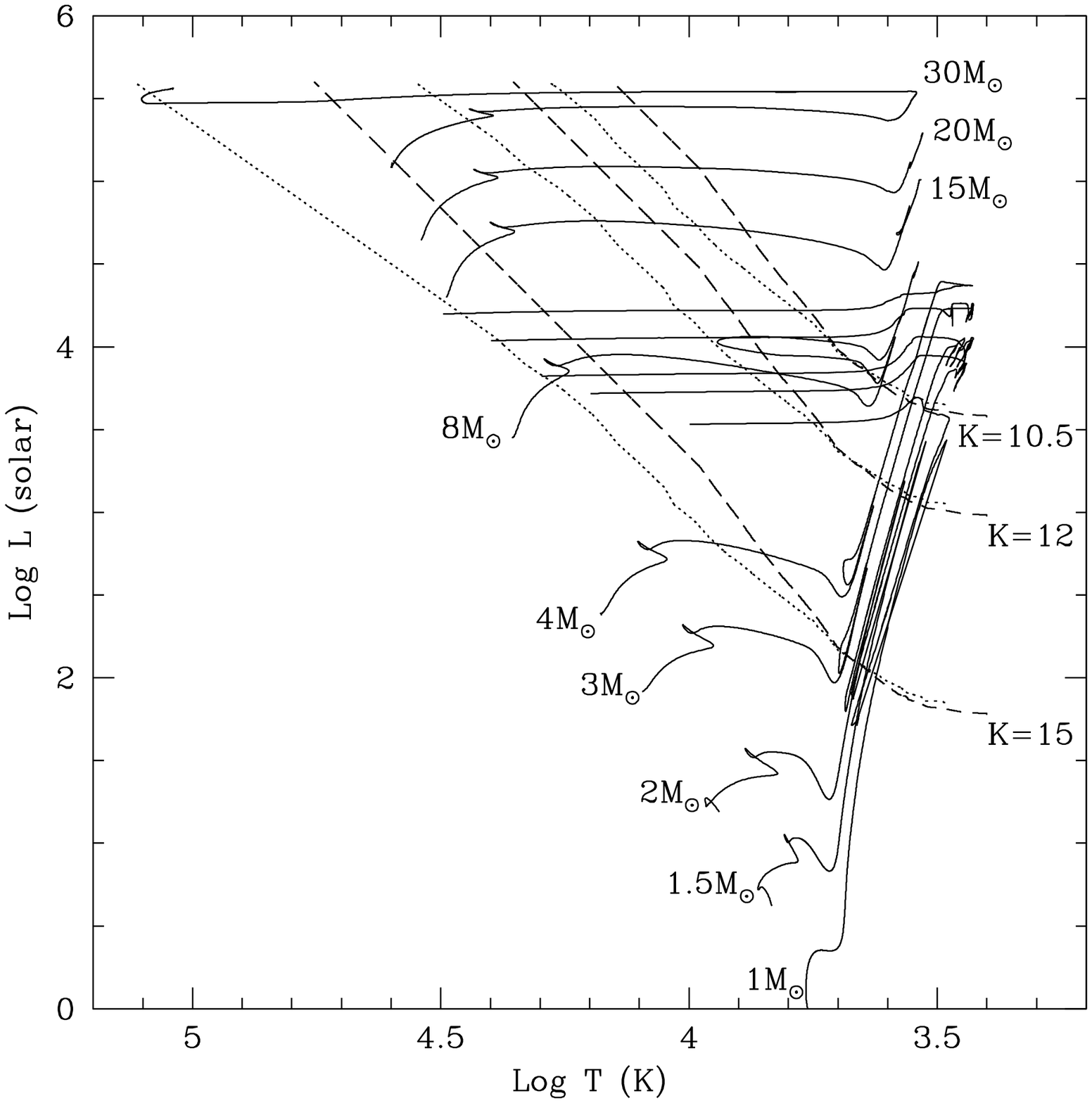}
\caption{Hertzprung--Russell diagram showing tracks of $1, 2, 3, 4, 8, 15, 20, 30$\,M$_{\odot}$ stars with contours of K=10.5, 12, 15 (corresponding to the boundaries of the bands observed in 
\protect \cite{1996ApJ...472..153G}) using main-sequence colours and bolometric corrections (dotted lines) and giant colours and bolometric corrections (dashed lines) from \protect \cite{1966ARA&A...4..193J}. A distance modulus of 14.6 \citep{1996ApJ...472..153G} and an extinction $A_{\mathrm{K}}$ of 3.0 \citep{1989ApJ...336..752R} have been assumed. Tracks are for solar--metallicity stars. Objects with $K<10.5$ are depleted within $\sim0.2$\,pc, objects with $10.5<K<12$ are depleted within $\sim0.08$\,pc and objects with $12<K<15$ are not depleted.}
\label{fig:hr_ktracks}
\end{figure}
Observations by \cite{1996ApJ...472..153G} in three different infra--red K--bands showed that there is a region very close to Sgr A$^{*}$ devoid of bright giant stars. They detected no late--type stars with K magnitudes brighter than $K=10.5$ within $\approx0.2$\,pc and no late--type stars with K magnitudes in the range $12>K>10.5$ within $\approx 0.08$\,pc of the supermassive black hole. In a fainter band, $15>K>12$, they observed that the surface density of late--type stars continued smoothly in to much smaller radii. For the remainder of the paper, we will refer to the three bands defined in \cite{1996ApJ...472..153G} as the bright, middle and faint bands respectively. Following \cite{1999MNRAS.308..257B} and referring to Figure 6 in \cite{1996ApJ...472..153G}, we estimate that the number of objects missing from the bright band is between eleven (assuming that the surface density distribution of these objects is flat inside 6 arcsec) and twenty--five (extrapolating the surface density inwards). Similarly, we estimate that the number of objects missing from the middle band is between six and eleven. The sharp drops in surface density at $\sim 6$ arcsec in the bright band and $\sim3$ arcsec in the middle band argue against the depletion being caused by peculiarities in the IMF and we do not think it can be due to tidal stripping, as will be explained in a later paper. Instead, the nature of the depletion strongly implies a collisional origin.\\
\subsection{Identifying the missing stars}
\begin{figure}
\includegraphics[width=0.45\textwidth]{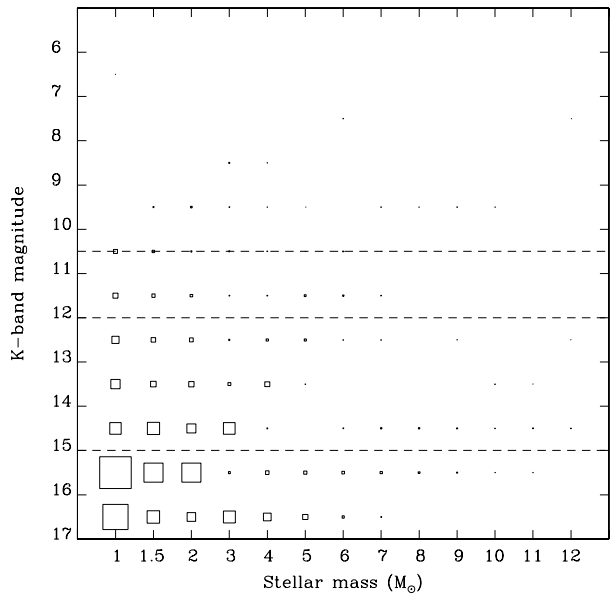}
\caption{Plot depicting the relative contributions of stars of mass $1-12$\,M$_{\odot}$ to K bands in the range $15-7$, as seen at the Galactic Centre. The area of the squares is proportional to the number of stars of each mass in each K--band assuming a Miller--Scalo IMF, a constant star formation rate over $14$\,Gyr and using the Eggleton stellar evolution tracks and colours and bolometric corrections from Johnson 1966. The boundaries of the K--bands observed by \protect \cite{1996ApJ...472..153G} are indicated by the dashed horizontal lines. The large concentration of objects with K=$15-16$ is the red clump.}
\label{fig:square_stars}
\end{figure}

\indent In Figure \ref{fig:hr_ktracks}, we show an HR diagram depicting the evolutionary tracks of $1, 2, 3, 4, 8, 15, 20$ and $30$M$_{\odot}$ stars derived using the STARS code \citep{1971MNRAS.151..351E, 1995MNRAS.274..964P}. We overlay K contours of $10.5, 12$ and $15$ magnitudes calculated using colours and bolometric corrections from \cite{1966ARA&A...4..193J} and adopting a distance modulus of 14.6 \citep{1996ApJ...472..153G} and an extinction $A_{\mathrm{K}}$ of 3.0 \citep{1989ApJ...336..752R} and assuming solar metallicity. Dotted lines show contours appropriate for main--sequence stars and dashed lines are contours appropriate for giants. This plot shows that objects brighter than $K=10.5$ are low--mass AGB stars or high--mass red supergiants, while those whose K magnitudes lie between $12$ and $10.5$ are likely to be low--mass red giants and AGB stars or high--mass main--sequence (MS) stars.\\
\indent To obtain a more detailed picture of which segment of the Galactic Centre stellar population is missing, we calculate the length of time spent by each star within a given K--band. If we assume that the stellar population in the Galactic Centre has been forming stars continuously at a steady rate following a Miller--Scalo IMF for 14 Gyr, we can estimate the relative proportions of objects in a population of this age which are visible at a given K magnitude. We also attempted realisations using different assumptions. We found that a Salpeter IMF was unable to reproduce the numbers of giants that \textit{are} observed outside the depleted regions in the Galactic Centre \citep[e.g.][]{1996ApJ...472..153G} in the K--bands using a realistic total mass for the Galactic Centre cluster -- the result is a giant population in which the bright band is overpopulated with respect to the middle and faint bands. We were unable to significantly improve the fit to the observed numbers of giants by varying the slope of the Salpeter mass function from its canonical value of 2.35 and the overabundance of bright giants becomes significantly worse for slopes flatter than 1.5. However, we note that, in accord with the Miller--Scalo IMF, Salpeter--like IMFs result in the bright K--band being largely populated by $2$--$4$M$_{\odot}$ objects and the middle band by $1$--$2$M$_{\odot}$ objects, unless the slope was $<1.5$. In addition, we found that assuming a burst of star formation (i.e. that the Galactic Centre stars are all the same age) was unable to reproduce the observed numbers of giants regardless of the time at which the burst occurred, since giants of a very narrow age (and therefore mass) range all have very similar K--band magnitudes. In particular, a single burst occurring $\gtrsim10$ Gyr ago would result in only giants of $\approx1$M$_{\odot}$ being visible now, leaving the bright band almost unpopulated throughout the Galactic Centre, which is not what is observed. We also attempted to reproduce the Galactic Centre cluster using an exponentially declining star formation rate, starting from a maximum at 14 Gyr. We found that, if the timescale on which the star formation rate drops off is significantly less than the assumed age of the cluster (14 Gyr), the model produces too few bright giants and too many faint ones. We concluded that, if the star formation rate has been declining, it must have done so slowly, so that the assumption of a constant rate is reasonable.\\
\indent In summary, the simple and conservative assumptions we have made about the IMF and star formation history of the Galactic Centre generate a model which matches the observed numbers of giants outside the depleted regions well and the alternative models we studied produced significantly worse fits. In Figure \ref{fig:square_stars}, we show the results of a Monte Carlo realisation of the Galactic Centre population constructed using these assumptions. The area of the square at each point in the (mass--K--magnitude) grid represents the relative number of stars of that mass which will be visible at that K magnitude. We see that the three bands observed by \cite{1996ApJ...472..153G} are dominated by low--mass objects in the range $1-4$\,M$_{\odot}$. There is a small contribution from higher--mass objects, but it is suppressed both by the IMF, which makes these objects intrinsically rare, and by their rapid evolution. Middle band objects ($12<K<10.5$) are likely to be $1-2$\,M$_{\odot}$ stars, while bright band objects ($K<10.5$) are likely to be $2-3$\,M$_{\odot}$ stars. Figures \ref{fig:hr_ktracks} and \ref{fig:square_stars} clearly show that the missing objects are giant stars with masses in the range $1-3$\,M$_{\odot}$ (the turnoff mass in our models is $\sim0.9$\,M$_{\odot}$). We now seek to explain why there should be a deficit of these stars near Sgr A$^{*}$.\\
\subsection{Possible causes of the depletion of giants}
\indent It has been suggested \citep[e.g.][]{2005ApJ...624L..25D} that stars on the RGB or AGB on eccentric orbits about Sgr A* could be tidally stripped of their envelopes as they pass through periapse. Since objects on eccentric orbits spend most of their time at apapse, this could potentially produce a depletion of such stars at distances from Sgr A* much larger than the separations required to tidally strip RGB or AGB stars. We think that tidal stripping of this nature will make at most a minor contribution to the depletion of giant stars in the Galactic Centre, since stripping must occur during the short RGB phase. This will form the subject of a later paper.\\
\indent Alternatively, numerous authors \citep{1996ApJ...472..153G, 1999MNRAS.308..257B, 1999ApJ...527..835A} have suggested that the depletion of red giants is due to collisions of the giants with other members of the Galactic Centre stellar population. Collisions would remove mass from the giants' envelopes and potentially alter their subsequent evolution so that they do not become bright enough to be visible in the depleted K bands. Since the Galactic Centre cluster is very densely--populated and has a high velocity dispersion ($\gtrsim100$\,km\,s$^{-1}$), collisions provide an attractive explanation for the lack of bright giants within the central $\sim0.1$\,pc.\\
\indent Previous work \citep[][]{1998MNRAS.301..745D,1999MNRAS.308..257B,1999ApJ...527..835A} has generally focussed on collisions impacting stars while they are on the RGB, and we largely agree that this is where the focus should be. It is highly likely that collisions of low mass MS stars with compact objects could expel enough mass to take the stars below the turnoff mass ($\sim0.9$\,M$_{\odot}$), ensuring that they never evolve to become giants within a Hubble time. However, this cannot be the sole cause of the depletion of giants in the Galactic Centre, as some giants \textit{are} observed there. In addition, the effects of collisions on a dense cluster of main sequence stars are complex, since most such collisions are likely to be with other MS stars. Some fraction of these encounters will lead to mergers resulting in the conversion of two stars to a single higher--mass (and therefore brighter, but also more shortlived) star. This coagulation process \citep[e.g.][]{1993ApJ...418..147L} and the effect on MS stars of collisions with stellar--mass black holes will form the subject of a companion paper.\\
\indent As will be explained in more detail in a later section, we have modelled the effect of mass loss on stars at various different stages in their evolution using the STARS code. The mass losses required to substantially alter the evolution of HB or AGB stars are very large -- $>99\%$ of the envelope -- and extremely unlikely even with a massive compact impactor. In addition, if a star is allowed to survive until the HB or AGB, it will already have been visible in the depleted K--bands for some of the duration of its RGB phase. Collisions occurring after the RGB will have very little effect on the observed stellar population; to alter the evolution of post--MS star so as to have a significant impact on the visible population, collisions must occur \textit{on the RGB}.\\
\indent \cite{1999MNRAS.308..257B} considered collisions between $2$\,M$_{\odot}$ and $8$\,M$_{\odot}$ giants and impactors of a variety of masses intended to represent MS stars, neutron stars and white dwarfs. For each giant--impactor pair, they explored the ($v_{\infty},R_{\mathrm{min}}$) parameter space relevant for the Galactic Centre. Throughout this paper, $v_{\infty}$ denotes the relative velocity between the colliding objects at infinite separation and $R_{\mathrm{min}}$ is the minimum distance that would be reached by their centres of mass, were they point masses on Keplerian hyperbolic orbits.\\
\indent Although \cite{1999MNRAS.308..257B} did not quantify how much mass either of their two giants must lose to prevent them evolving to become visible in the middle or bright bands, they observed that the mass losses resulting from encounters of giants with MS stars, white dwarfs and neutron stars were not large, of the order of a few to ten percent. They concluded that such collisions were not able to explain the observed depletion of giants.\\
\indent In this paper, we extend the work of \cite{1999MNRAS.308..257B} by considering, in addition to impacts with MS stars, encounters between giants and stellar--mass black holes. We also consider the effects of multiple impacts involving a single giant. We examine the effects of collisions involving $1-3$M$_{\odot}$ giants and black holes or MS stars.\\
\section{Collision rates in the Galactic Centre}
\begin{figure}
\includegraphics[width=0.45\textwidth]{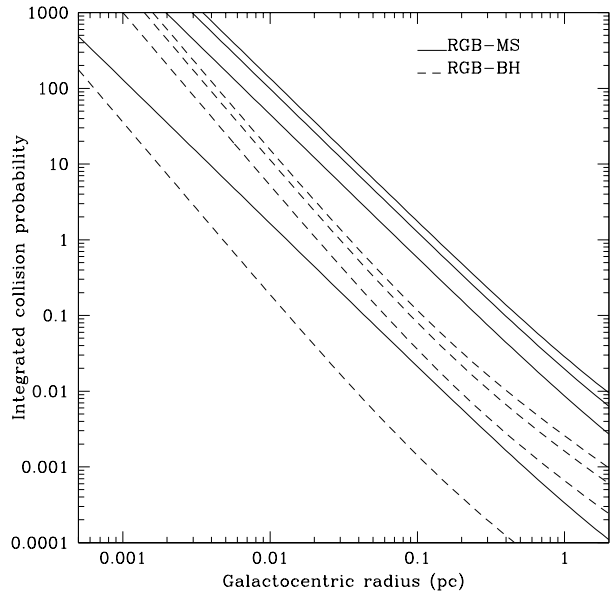}
\caption{Plot showing the integrated probability that stars of masses (from top to bottom) $1$, $1.5$, $2$ and $3$\,M$_{\odot}$ will collide with main--sequence stars (solid lines) or stellar--mass black holes (dashed lines) while themselves on the red giant branch, as a function of Galactocentric radius.}
\label{fig:coll_rates}
\end{figure}

Within a few pc of the Galactic Centre, the stellar distribution follows a broken power--law cusp \citep{2003ApJ...594..812G, 2007A&A...469..125S}. \cite{2003ApJ...594..812G} find that the three--dimensional number density of visible stars $n_{*}$ varies with Galactocentric radius as
\begin{eqnarray}
n_{*}(r)\propto\left\{\begin{array}{lll}
r^{-1.4} & \textrm{for} & r\leq 0.4\textrm{pc}\\
r^{-2.0} & \textrm{for} & r>0.4\textrm{pc} 
\end{array} \right.
\end{eqnarray}
and \cite{2007A&A...469..125S} find
\begin{eqnarray}
n_{*}(r)\propto\left\{\begin{array}{lll}
r^{-1.2} & \textrm{for} & r\leq 0.22\textrm{pc}\\
r^{-1.75} & \textrm{for} & r>0.22\textrm{pc} 
\end{array} \right.
\end{eqnarray}
Since the break radii of both of these power--law models comfortably exceed the radius within which depletion of bright giants is observed ($\simeq0.1$\,pc), we adopt a single power--law model for the visible stars of the form
\begin{eqnarray}
n_{*}(r)\propto r^{-1.4}
\label{eqn:nms_r}
\end{eqnarray}
Strictly adopting the power law of \cite{2007A&A...469..125S} affects only slightly the rates of collisions between giants and MS stars, and has little influence on our results.\\
\indent Normalising Equation \ref{eqn:nms_r} at Galactocentric radii of $\sim0.1$\,pc is difficult, since the enclosed mass at this radius is dominated by the SMBH. The influence radius $r_{\mathrm{infl}}$, defined as the radius within which the enclosed stellar mass is equal to the mass of Sgr A* is of the order of $1$\,pc \citep[e.g][]{2006ApJ...649...91F}.\\
\indent To obtain a realistic normalisation at small radii, we use the results of simulations of the Galactic Centre performed by \cite{2006ApJ...649...91F} using the ME(SSY)**2 Monte Carlo stellar dynamics code \citep{2001A&A...375..711F, 2002A&A...394..345F}. The initial conditions were $\eta$--models \citep{1993MNRAS.265..250D} with $\eta=1.5$ and a central massive object representing Sgr A* (models with $\eta=2$, corresponding to a shallower cusp, were also run, but produced very similar results). The mass of the central object, the mass of the Galactic Centre cluster and the break radius at which the $\eta$--model steepens to $\rho(R)\propto R^{-4}$ were taken to be $3.5\times10^{6}$\,M$_{\odot}$, $7\times10^{7}$\,M$_{\odot}$ and $14$\,pc respectively, designed to match observations of the Galactic Centre by \cite{2003ApJ...596.1015S} and \cite{2005ApJ...620..744G}. \cite{2006ApJ...649...91F} found that the MS stars relax into a cusp with $\gamma=1.3-1.4$, in agreement with the observations of \cite{2003ApJ...594..812G}. Using their results (e.g. their Figure 2), we adopt a normalisation for the main sequence stars such that $M(0.1\mathrm{pc})=4\times10^{4}$\,M$_{\odot}$. For an evolved stellar population with a Miller--Scalo IMF, the mean stellar mass is $\sim0.5$\,M$_{\odot}$, so this normalisation corresponds to $8\times10^{4}$ main--sequence stars within a Galactocentric radius of $0.1$\,pc.\\
\indent The population of stellar--mass black holes in the Galactic Centre is expected to have the form of a cusp with a power--law exponent of $1.75$ provided that they dominate the relaxation rate \citep{1976ApJ...209..214B}. This relation should hold at least in the innermost few tenths of a parsec \citep[][and references therein]{2007astro.ph..3495A}. However, setting the normalisation of this population is very difficult, since the black holes are not directly observable.\\
\indent \cite{1993ApJ...408..496M} estimated that all the black holes formed within $\sim4$\,pc of the Galactic Centre would sink into the centre over the age of the Galaxy, resulting in a total mass in stellar black holes there of $\sim10^{6}$\,M$_{\odot}$. They were not able to say whether such a migration would result in a dense cluster of black holes, or would simply add mass to Sgr A*. \cite{2000ApJ...545..847M} estimated, in good agreement with \cite{1993ApJ...408..496M}, that all black holes formed within $5$pc would migrate in over the age of the Bulge. Using a model of the stellar mass density as a function of Galactocentric radius from \cite{2000MNRAS.317..348G} and assuming a piecewise IMF from which all stars in the mass range $30-100$\,M$_{\odot}$ yield black holes, they estimated that $2.4\times10^{4}$ BHs should have migrated into the central $0.7$\,pc. They calculated that the timescale on which stellar--mass BHs would be captured by Sgr A* would be considerably longer than a Hubble time, so that all these stellar--mass BHs should still be within this volume. A similar indirect means of estimating the number of black holes within a given Galactocentric radius is given by \cite{2004ApJ...606L..21A}. They assume that stellar--mass BHs have migrated into the Galactic Centre by mass segregation and estimate the maximum number that can survive there by requiring that the timescale on which the BH population is lost by scattering into the loss cone is longer than a Hubble time. For a population of $10$\,M$_{\odot}$ BHs, their analysis yields a maximum number of BHs that can be within $0.1$\,pc of Sgr A$^{*}$ of $\sim5 000$.\\
\indent Stellar--mass black holes can be directly observed if they are accreting material at sufficient rates to be visible as X--ray sources. \cite{2005ApJ...622L.113M} reported four X--ray transients within $1$\,pc of Sgr A$^{*}$, an overabundance of a factor $\sim20$ when compared to the field. \cite{2003ApJ...591..891B} detected diffuse X--ray emission with a $2-10$\,keV luminosity of $10^{33}$\,erg\,s$^{-1}$ coincident with the position of Sgr A$^{*}$ to within $0.27$\,arcsec and with an intrinsic size of $0.61$\,arcsec. They consider the possibility that the source may be a population of compact objects accreting from the ambient gas, but point out that the luminosity of such a system would be orders of magnitude less than that of Sgr A$^{*}$ itself accreting from the same gas. They point out that a single X--ray binary, in which the compact object is supplied with a strong accretion flow, would have a luminosity comparable to that observed. However, the number of X--ray binaries expected within this volume ($r<0.03$pc) is very small, since binaries are difficult to form in a region with such a high velocity dispersion, and would also be quickly disrupted by stellar encounters. Taking into account that the observed extent of the source is consistent with Sgr A$^{*}$'s Bondi radius, \cite{2003ApJ...591..891B} conclude the emission is most likely to be due to accretion onto the supermassive black hole itself.\\
\indent \cite{2007MNRAS.377..897D} examined the question of whether the X--ray emission could be due to compact objects accreting from the ISM, specifically the Minispiral \citep{2004A&A...426...81P}, in more detail. They modelled the Minispiral as a half--disc of $50$\,M$_{\odot}$ of gas in a circular orbit between $0.1$ and $0.5$\,pc around Sgr A$^{*}$. They constructed a Bahcall and Wolf cusp of black holes extending to $0.7$\,pc, allowed the black holes to accrete the gas and calculated their instantaneous luminosity as they moved along their orbital tracks. By following how many black holes were detectable at any given time, they placed constraints on the maximum number of compact objects. They found that the number of stellar--mass BH within $0.1$\,pc was likely to be at most a few thousand. None of these authors consider the possibility that X--ray sources may be lone stellar--mass black holes accreting material captured during a stellar collision.\\
\indent We again use the results of \cite{2006ApJ...649...91F} to obtain a fiducial model for the number and distribution of black holes in the innermost few tenths of a parsec. \cite{2006ApJ...649...91F} found that the stellar mass black holes migrated inwards by mass segregation, forming a cusp with $\gamma=1.8$, in very good agreement with \cite{1976ApJ...209..214B}, and that the central $0.1$\,pc contained a few$\times10^{3}$ BH. We therefore adopt a black hole population with a cusp of the form 
\begin{eqnarray}
n_{\rm BH}(r)\propto r^{-1.8},
\label{eqn:nbh_r}
\end{eqnarray}
normalised so that $n_{\rm BH}(0.1\mathrm{pc})=5\times10^{3}$. The radius at which the total enclosed mass of our MS and BH populations becomes equal to the mass of Sgr A* ($4\times10^{6}$\,M$_{\odot}$) is then $\sim1.4$\,pc\\
\indent In this work, it will be assumed that stars lie inside the sphere of influence and hence that their orbits are Keplerian with one--dimensional velocity dispersion $\sigma_{\rm 1D}$ is $\propto r^{-1/2}$. If the stellar density distribution has the form of a cusp with three dimensional radial distribution $\propto r^{-\gamma}$, then the one--dimensional velocity dispersion is given by
\begin{eqnarray}
\sigma_{\rm 1D}(r)=\left[\frac{GM}{(\gamma+1)r}\right]^{\frac{1}{2}}.
\label{eqn:v_rel}
\end{eqnarray}
\indent If a star of mass $M_{*}$ and radius $R_{*}$ moves through a cluster of impactors of mass $M_{\mathrm{imp}}$ and radius $R_{\mathrm{imp}}$ such that the relative encounter velocity is $v_{\infty}$, the effective cross section $s$ of the star is given by
\begin{eqnarray}
s=\pi (R_{*}+R_{\rm imp})^{2}\left\{1+\frac{2G(M_{*}+M_{\rm imp})}{(R_{*}+R_{\rm imp})v_{\infty}^{2}}\right\}.
\label{eqn:cross_section}
\end{eqnarray}
The second term inside the braces on the right--hand side of Equation \ref{eqn:cross_section} is the increase in the star's effective cross section due to gravitational focussing.\\
\indent At a given Galactocentric radius $r$, stars will in reality have a distribution of velocities which must be integrated over to yield an accurate determination of the collision rate at that radius. We find that the velocity distribution in an $\eta$--model is very similar to a Maxwellian distribution. Following \cite{1987gady.book.....B}, we define the collision time averaged over the velocity distribution by
\begin{eqnarray}
\begin{array}{lll}
t_{\rm col}(r)^{-1}=&4\sqrt{\pi}(R_{*}+R_{\rm imp})^{2}n_{\rm imp}(r)\sigma_{\rm 1D}(r)\times& \\
&&\\
&\left[1+\frac{G(M_{*}+M_{\rm imp})}{2\sigma_{\rm 1D}^{2}(R_{*}+R_{\rm imp})}\right] & 
\end{array} 
\end{eqnarray}
where $n_{\rm imp}(r)$ is the number density of the impactors.\\
\indent Since the radius of a star changes drastically during its lifetime, $R_{*}=R_{*}(t)$ and $t_{\mathrm{col}}$ is a function of time as well as Galactocentric radius. As a result, a more useful quantity is the integrated probability $p(r)$ of the star experiencing a collision over its lifetime, given by
\begin{eqnarray}
p(r)=\int_{0}^{\tau_{*}}\frac{dt}{t_{\rm col}(r,t)}
\label{fig:pcol}
\end{eqnarray}
where $\tau_{*}$ is the lifetime of the star. In this paper, we are primarily interested in collisions occurring to stars while they are on the giant branch, in which case the collision probability of interest is
\begin{eqnarray}
p_{{\rm RGB}}(r)=\int_{t({\rm BRGB})}^{t({\rm TRGB})}\frac{dt}{t_{\rm col}(r,t)}
\label{eqn:pcol_rgb}
\end{eqnarray}
where $t(\mathrm{BRGB})$ is the age of the star at the base of the red--giant branch and $t(\mathrm{TRGB})$ is its age at the tip. We define the base of the red--giant branch as the point at which the star's luminosity reaches a minimum at the end of the Hertzprung gap, and the tip of the red--giant branch as the point where the star's luminosity reaches its maximum value on the giant branch.\\
\indent In Figure \ref{fig:coll_rates}, we plot the integrated collision probability given by Equation \ref{eqn:pcol_rgb} as a function of Galactocentric radius for $1$, $1.5$ $2$ and $3$\,M$_{\odot}$ stars, assuming that the impactors are $1$\,M$_{\odot}$ main--sequence stars (solid lines) or $10$\,M$_{\odot}$ black holes, distributed according to Equations \ref{eqn:nms_r} and \ref{eqn:nbh_r}. For reasons of clarity, we have assumed that the inner power law $\alpha=1.4$ holds at all radii -- since the break in the  cusp power law derived by \cite{2003ApJ...594..812G} occurs at radii where the integrated probability of suffering a collision with a MS impactor is small ($<10\%$), this does not affect our results.\\
\indent In deriving Equation \ref{eqn:pcol_rgb}, it is assumed that $n_{\mathrm{imp}}(r)$ and $v_{\infty}$ are constant over the lifetime of the star, which can only be true if either the star's orbit is circular, or if the star's lifetime is much shorter than its orbital period. Since neither of these of these conditions is true in general, we integrated the collision probability self--consistently on orbits with a variety of eccentricities. The effect of orbital eccentricity was to increase the probability of a star with a given semimajor axis suffering a collision by factor of at most $2$ over the collision probability obtaining from a circular orbit with the same semimajor axis.\\
\indent A further simplification made in estimating these collision rates is that the impactors are all either $1$\,M$_{\odot}$ main--sequence stars or $10$\,M$_{\odot}$ black holes. Given the lack of knowledge about the mass function of stellar--mass black holes, the latter assumption is reasonable (and will be discussed in more detail later), but the former may not be. We therefore recalculated the collision rates assuming that the main--sequence impactors were distributed according to a Miller--Scalo IMF between $0.08$ and $1$M\,$_{\odot}$. We found that this makes little difference to the collision rates.\\
\indent We see from Figure \ref{fig:coll_rates} that the probability of a $1-2$\,M$_{\odot}$ giant suffering a collision with a main--sequence star exceeds unity within a radius of $\sim0.1$\,pc and the probabilities of the same giants being struck by a black hole exceeds unity within a radius of $\sim0.03$\,pc. The $3$\,M$_{\odot}$ star has a very short giant phase in comparison to the lower--mass objects, and the radii at which the probabilities of the $3$\,M$_{\odot}$ giant being struck approach unity are approximately an order of magnitude smaller.\\
\indent We interpret collision probabilities in excess of unity as implying that stars at that radius will suffer more than one collision. Figure \ref{fig:coll_rates} shows that the region of space within $0.1$\,pc of Sgr A$^{*}$ is a highly collisional environment. In the following sections, we investigate the effect of stellar collisions on the population of bright giant stars in the Galactic Centre.\\
\section{Numerical methods}
\subsection{Stellar evolution code}
\indent To evolve the stellar models presented in this paper we used STARS, the
Cambridge Stellar Evolution code.  STARS was originally written by Peter
\citet{1971MNRAS.151..351E} and has been extensively modified since (see
\citealt{1995MNRAS.274..964P} and references therein for a complete
description).  We use the Reimers mass-loss law on the RGB
\citep{1978A&A....70..227K} with $\eta=0.4$ and the
\citet{1993ApJ...413..641V} mass-loss law on the AGB; on the MS the
mass-loss is assumed to be negligible for the low-mass stars considered
here.  Convective overshooting was included with $\delta_{\rm ov}=0.12$
\citep{1997MNRAS.285..696S}.\\
\indent Following \cite{2006A&A...452..295E} we modify our mass-loss rates for
high-mass stars.  For OB stars, we use the mass-loss predictions of \cite{2001A&A...369..574V} which scale with metallicity as $(Z/Z_\odot)^{1/2}$.  For all
other pre-WR phases, we employ the rates of \cite{1988A&AS...72..259D} scaled
similarly with metallicity. When the star becomes a WR star ($X_{\rm surf}
< 0.4$ and $\log(T/{\rm K}) > 4.0$), we use the rates of \cite{2000A&A...360..227N}.
\subsection{Smoothed particle hydrodynamics}
For the hydrodynamical simulations presented here, we used a Smoothed Particle Hydrodynamics (SPH) code based on that described in \cite{1990nmns.work..269B}. The code uses the SPH formalism to solve fluid equations and a binary tree to calculate gravitational forces. The code has been modified to include point masses, particles which only interact with other particles via gravitational forces and are used to model compact objects \citep{1991ApJ...381..449D}. Their gravitational fields are smoothed in the same way as those of ordinary SPH particles, although the smoothing lengths of the point masses are fixed. We use the standard SPH artificial viscosity formalism with $\alpha=1$, $\beta=2$. Gas particles are assumed to behave adiabatically.\\
\indent We constructed SPH stellar models using one--dimensional density and temperature profiles from the STARS code. SPH particles are placed on a uniform hexagonal close--packed grid and their masses iteratively adjusted so that the local density matched that of the one--dimensional model.\\
\indent The extreme density gradients in the envelopes of giant stars, both very near the core and near the stellar surface, are difficult to model. In order to reproduce the density profiles of the outer envelopes of our giants while retaining sufficient resolution of the material near the core and avoiding very large particle numbers, we deformed the uniform particle grid in our models. Using results from our stellar evolution calculations, we determined the critical mass loss required to prevent the giant evolving to become bright enough to appear in the depleted K--bands. This gave us a minimum mass (enclosed by a radius $r_{crit}$) that we needed to resolve. We constructed a uniform particle grid so that $r_{crit}$ contained at least $2000$ particles. The particle grid was then deformed according to\\
\begin{eqnarray}
r_{\rm f}=\left\{\begin{array}{lll}
r_{\rm i}\mathrm{cosh}\left[\alpha\left(\frac{r_{\rm i}}{r_{\rm crit}}-1\right)\right] & \textrm{for} & r_{\rm i}>r_{\rm crit} \\
r_{i} & \textrm{for} & r_{\rm i}\le r_{\rm crit}
\end{array}\right.
\end{eqnarray}
where $r_{\rm i}$ is the initial radius of a given particle, $r_{\rm f}$ is the final radius and $\alpha$ is a constant ensuring the outer radius of the deformed grid is equal to the radius of the star.\\
\indent We modelled the giant cores as point masses. To stabilise the interaction between the point mass core and the envelope, the innermost 477 gas particles (i.e. those surrounding the core, corresponding to an enclosed mass about $1\%$ of the envelope) were `frozen' -- their velocities were constrained during simulations to be equal to that of the core particle. Ensuring adequate resolution of the outer envelope and the core region and guaranteeing that the model be stable required $\sim 175,000$ particles.\\
\indent The models were allowed to relax in isolation for several freefall times with a linear damping term acting to remove any residual oscillations in the particle grids. Collision simulations were started with the stars sufficiently far apart that tidal effects were negligible.\\
\indent Most calculations were run on single cpus on PCs. However, in order to examine numerical convergence, we repeated several simulations at high--resolution on the UK Astrophysical Fluids Facility at Leicester University, UK. In these calculations, we used $\sim1.2$ million particles in our giant models. The factor of $\sim7$ greater particle number gives a factor of $\sim2$ better linear resolution.\\
\subsection{Restricted three--body code}
\indent In order to extend the range of simulations we could perform without consuming unacceptable quantities of computer time, and as an independent check on our SPH calculations, we also studied red giant collisions using a simple restricted three body code. As pointed out by \cite{1988ApJ...332..271L}, if the impact velocity is much greater than the speed of sound in the giant's envelope, the envelope cannot react to the passage of the intruder. The interaction can then be approximated by an encounter between two point masses (the impactor and the giant core) with the envelope represented by a third, extended particle with the enclosed mass profile of the undisturbed envelope. The force between the core and the intruder $F_{\rm c-int}$ is then given by
\begin{eqnarray}
F_{\rm c-int}=\frac{GM_{\rm core}M_{\rm int}}{r_{\rm c-int}^{2}}
\end{eqnarray}
and the force between a point particle $i$ and the envelope particle is given by
\begin{eqnarray}
F_{\rm i-env}=\frac{GM_{\rm i}M_{\rm env}(r_{\rm i-env})}{r_{\rm i-env}^{2}}
\end{eqnarray}
where $M_{\rm env}(r_{\rm i-env})$ is the envelope mass enclosed within a radius $r_{\rm i-env}$. The equations of motion of the three particles can then be easily integrated.\\
\indent In collisions in which the impactor imparts a significant velocity to the core relative to the envelope, the fraction of the envelope retained by the core can be estimated by computing the mass within the Bondi--Hoyle radius, given by
\begin{eqnarray}
r_{\rm BH}=\frac{GM_{\rm core}}{v_{\rm rel}^{2}}
\end{eqnarray}
where $v_{\rm rel}$ is the relative velocity between the core and the envelope. This analysis can be improved upon slightly by instead computing the mass contained within the Bondi--Hoyle--Lyttleton radius, given by
\begin{eqnarray}
r_{\rm BHL}=\frac{GM_{\rm core}}{v_{\rm rel}^{2}+c_{\rm s}(r_{\rm BHL})^{2}}
\end{eqnarray}
where $r_{\rm BHL}$ must be found iteratively since it depends on the sound speed at that radius. We find in practice that the two schemes give very similar results.\\
\indent The envelope mass retained by the core may thus be estimated for any given  collision by using the three--body technique to find the relative velocity induced between the core and the envelope, yielding $r_{\rm BH}$ or $r_{\rm BHL}$, and determining the enclosed mass at this radius from the giant's density profile.\\
\subsection{Monte Carlo multiple--collision code}
\indent Since the collision rates shown in Figure \ref{fig:coll_rates} imply that giants may suffer multiple collisions with a mixture of black--hole and main--sequence impactors, we explored this possibility using a Monte Carlo technique. For a population of $10^{4}$ giants, we simulated collisions at all the relative velocities for which we have SPH simulations ($400$, $800$ and $1200$\,km\,s$^{-1}$ for the $1$ and $2$\,M$_{\odot}$ giants, $800$\,km\,s$^{-1}$ only for the $1.5$ and $3$\,M$_{\odot}$ objects) and at randomly--chosen periastrons (distributed so that the probability of a collision at a periastron $R$ is proportional to $R^{2}$, appropriate for non--focussed encounters) with the numbers of black hole and main--sequence impactors being set by the collision rate, and thus the Galactocentric radius. For each impact, we first constructed a synthetic mass--loss--against--$R_{\mathrm{min}}$ curve from the results of the SPH and three--body simulations of that encounter. As we will show in detail later, the three--body code agrees very well with the SPH calculations in encounters at small periastrons, in which the core is ejected from the giant, but hydrodynamic effects become important at larger periastrons (particularly at lower relative velocities). We therefore use the smooth three--body curve for very close encounters where our SPH calculations are sparse, out to a periastron where it begins to underestimate the mass loss derived from the SPH calculations, and from this periastron outwards, we use the SPH mass--loss curve. Using this synthetic mass--loss curve, we calculate the mass lost at the randomly--chosen periastron by interpolation. The mass lost in successive collisions can either be assumed to be multiplicative or additive. If a given collision removes a fraction of mass $f_{\mathrm{i}}$, the total mass lost $f_{\mathrm{tot}}$ after $N$ collisions can be estimated by $1-f_{\mathrm{tot}}=(1-f_{1})(1-f_{2})...(1-f_{N})$, or by $f_{\mathrm{tot}}=f_{1}+f_{2}+$...$+f_{N}$. We found that it made little difference to the results which of these assumptions was made. From the total mass lost by each giant, we determined the fraction of the $10^{4}$ objects prevented from evolving into the bright and middle bands for each impactor number.\\
\section{Modelling the effect of collisional mass loss on the evolution of stars}
\begin{figure*}
\includegraphics[width=\textwidth]{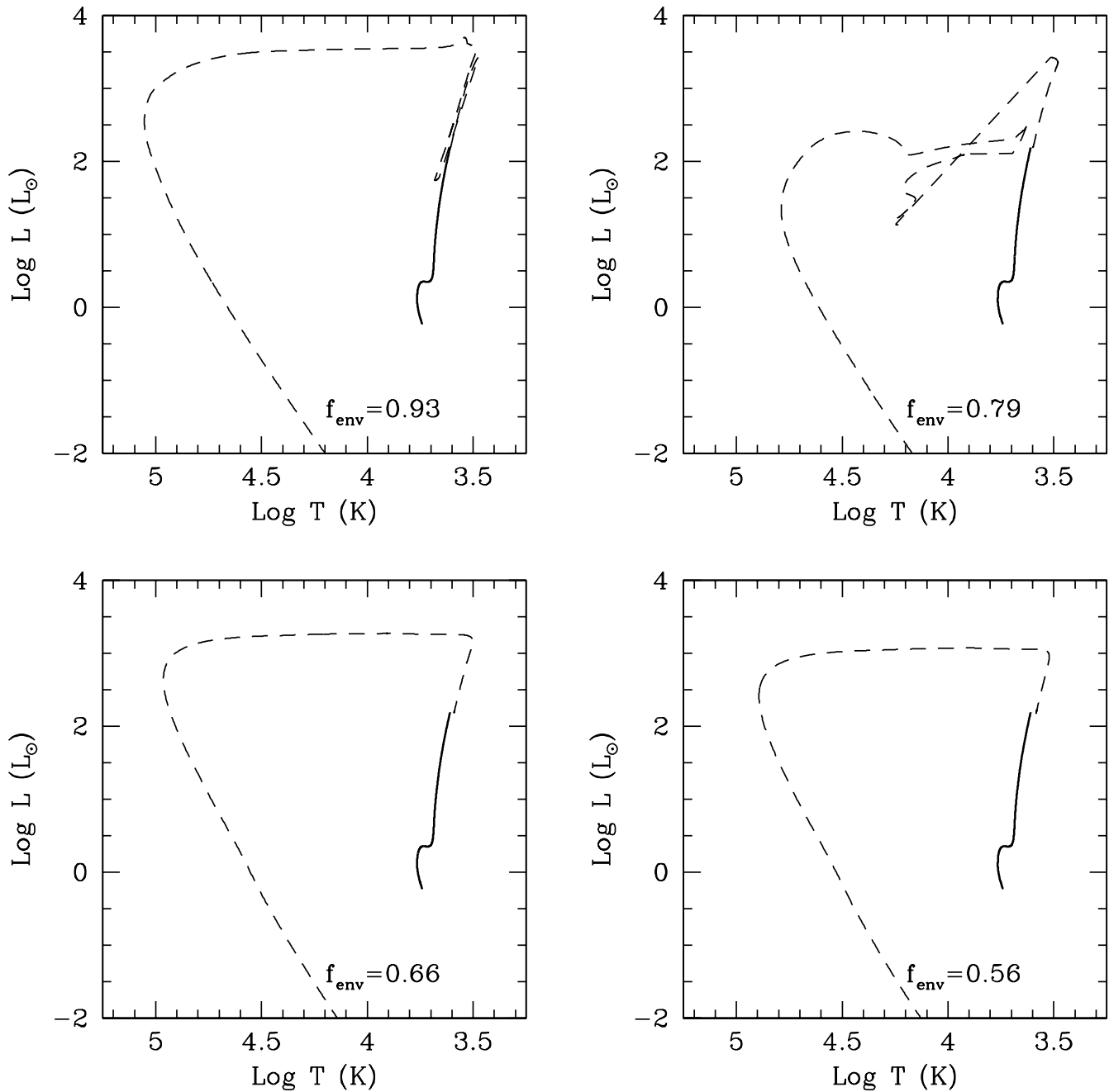}
\caption{Plot showing the effect on the evolution of a $1$M$_{\odot}$ star of removing increasing fractions of its envelope mass, once it has reached a point halfway up the RGB. The star's normal evolution up to this point is shown by the thick black line, and the evolution after mass loss by the dashed line. The fraction of the object's envelope remaining is given in the bottom right of each panel.}
\label{fig:hr_compare}
\end{figure*}

\begin{figure*}
\includegraphics[width=\textwidth]{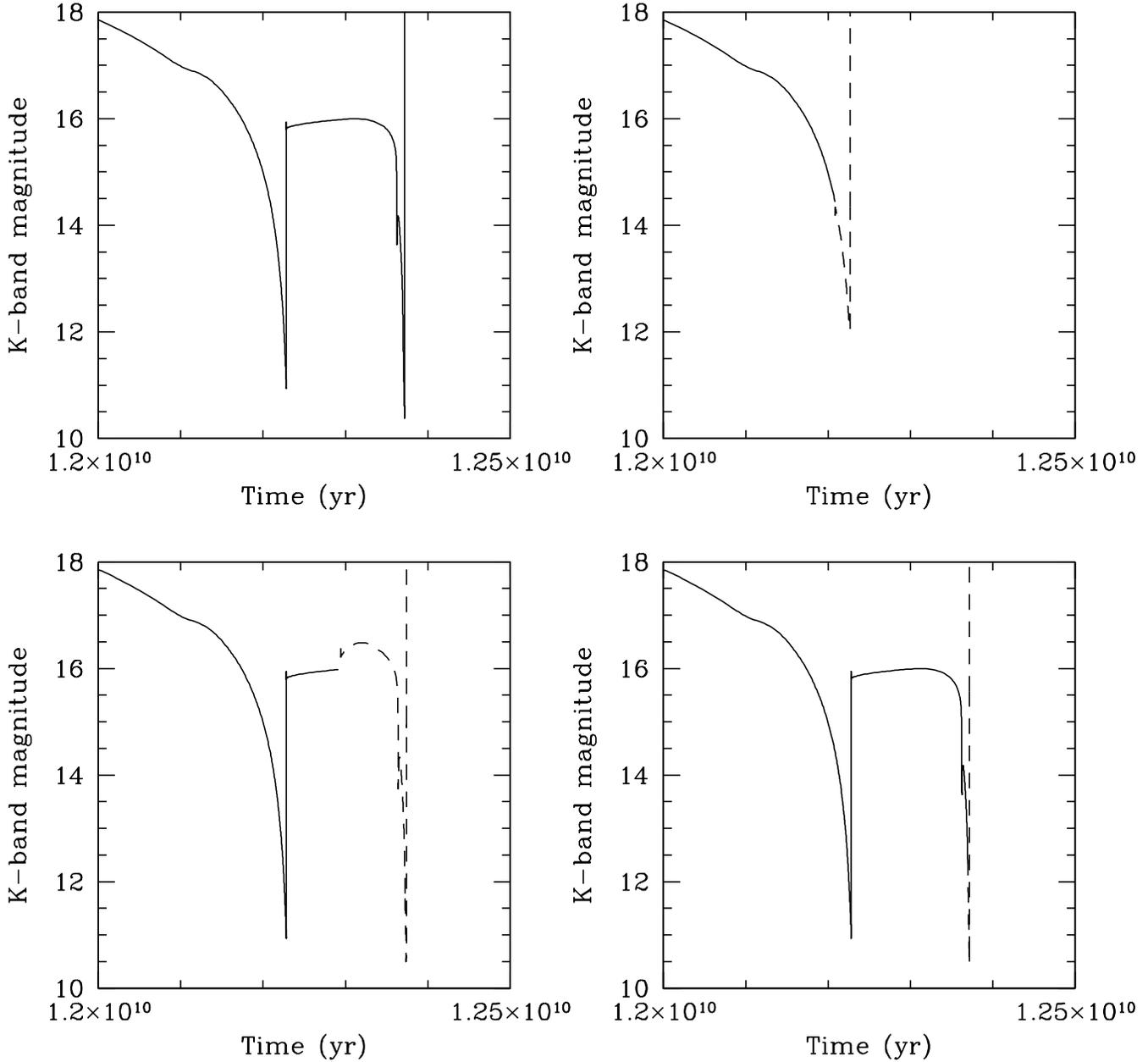}
\caption{Plot showing the effect on the evolution of the K--band brightness of a $1$\,M$_{\odot}$ star of removing $60\%$ of its envelope mass halfway along the RGB (top right panel), halfway along the HB (bottom left panel) and on the early AGB (bottom right panel). Top left panel and solid lines in other plots represent undisturbed evolution.}
\label{fig:HB_AGB_mass loss}
\end{figure*}

\begin{figure}
\includegraphics[width=0.45\textwidth]{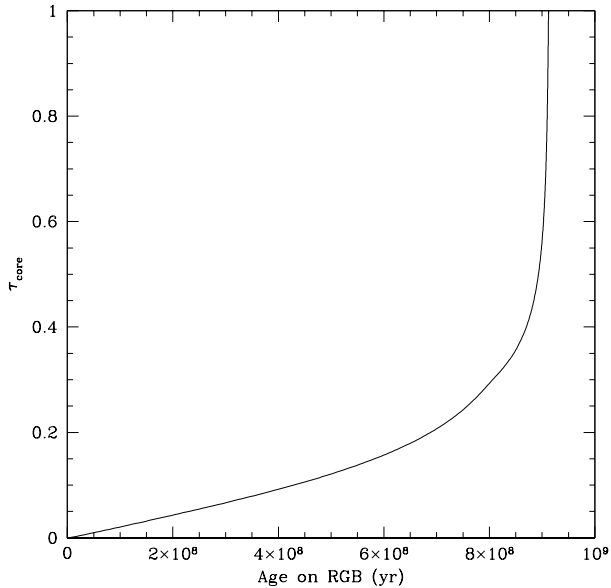}
\caption{Plot showing the relationship between $\tau_{\mathrm{core}}$ and elapsed time on the RGB for a $1$\,M$_{\odot}$ giant.}
\label{fig:clobberco_realtime}
\end{figure}

\begin{figure}
\includegraphics[width=0.45\textwidth]{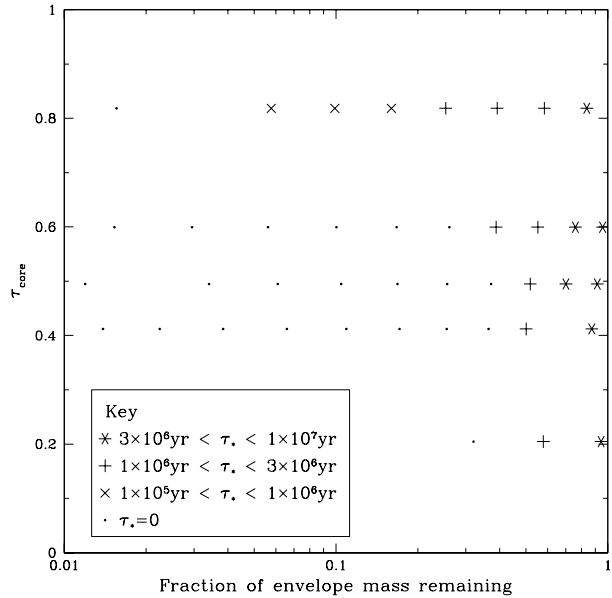}
\caption{Plot showing the effect on the timescale for which a $1$\,M$_{\odot}$ giant is visible in the middle band ($12>$K$>10.5$) of stripping mass from the giant at various stages of evolution on the giant branch. The evolutionary stage at which mass is stripped is denoted by the increase in core mass as a fraction of the total change in the core mass along the whole giant branch (defined in Equation \ref{eqn:clobber_coordinate}), given on the $y$--axis and the fraction of the envelope mass remaining after stripping is given on the $x$--axis. The symbols denote the length of time for which each stripped model is visible in the middle band for the duration of its life.}
\label{fig:giant_1msun_clobber}
\end{figure}

To study the effects of mass loss on the giants, we constructed in the STARS code models of stars of $1$, $1.5$, $2$ and $3$\,M$_{\odot}$ at several different ages on the main sequence, red giant branch, horizontal branch and AGB.\\ 
\indent In order to simulate the rapid mass loss caused by a collision we removed the desired quantity of mass from the star at a rate of
$5\times10^{-6}$M$_{\odot}{\rm yr}^{-1}$, which is the largest mass-loss rate at
which we could reliably converge models for low-mass stars. This rate is much lower than the mass--loss rates observed in our SPH calculations, which could reach $\sim10^{4}$M$_{\odot}$yr$^{-1}$ (the equivalent of $\sim1$M$_{\odot}$hr$^{-1}$)but is still much faster than the evolutionary timescales for low--mass stars. During this time
we turned off composition changes owing to nuclear burning but retained the
energy generation; this prevents nuclear evolution occurring whilst the mass
is removed but also keeps the star's structure consistent.  This is
equivalent to making the collisional mass-loss instantaneous on a nuclear
timescale.\\
\indent We removed increasing fractions of the models' envelopes and studied their evolution, using the \cite{1966ARA&A...4..193J} colours and bolometric corrections to calculate the K--band magnitude of the objects as functions of time. We were therefore able to determine, for a star of a given mass, how much mass a collision at any given stage in the star's life would need to expel such that the star would never evolve to become brighter than a given K--magnitude. In Figure \ref{fig:hr_compare}, we show by means of four HR diagrams the effect on the evolution of a $1$M$_{\odot}$ star of removing increasing fractions of its envelope at a point halfway up the giant branch. In each panel, the normal evolution of the star is denoted by the thick black line. The star is then taken to instantaneously lose mass, leaving it with the envelope fraction given in each figure panel. We see that leaving the object with $93\%$ of its envelope has little effect on its evolution -- it follows a normal AGB track and becomes a white dwarf. However, leaving the star with $79\%$ of its envelope results in an object resembling a subdwarf--B star, with a short AGB significantly fainter than that of the undisturbed star (although the luminosity of the tip of the RGB is not significantly changed). Leaving the star with $66\%$ or $56\%$ of its envelope, as shown in the lower panels of Figure \ref{fig:hr_compare} prevents the star igniting helium, thus removing the AGB phase altogether, and also makes the tip of the RGB fainter. In summary, removing a few tens of percent of the envelope of a low mass giant is sufficient to make its AGB phase significantly fainter, but to radically alter its evolution such that it never ignites helium at all requires the removal of $\sim40\%$ of the envelope.\\
\indent In all post--main--sequence phases, the fractions of the stars' envelopes that must be expelled in order to significantly affect their evolution are substantial, and especially so for HB or AGB stars. We find that, in order to significantly alter the evolution of a star, mass must be removed whilst it is on the main--sequence or the red giant branch. In Figure \ref{fig:HB_AGB_mass loss}, we illustrate this point by comparing the effect of removing $60\%$ of a $1$M$_{\odot}$ star's envelope at three different phases of its life. Each panel in the plot shows the evolution with time of the star's K--band magnitude. The top left panel shows the evolution of the undisturbed object. The top--right panel shows the result of removing $60\%$ of the envelope mass halfway along the RGB, the bottom left panel shows the result of removing this mass halfway along the horizontal branch and the bottom right panel shows the result of removing mass on the early AGB. In all cases, the solid lines represent evolution of the undisturbed object and dashed lines represent evolution after mass loss has occurred. The plots clearly show that this mass loss occurring on the RGB has a strong effect on the star's evolution, making the tip of the giant branch $\sim1.5$ magnitudes fainter (although having little effect on the duration of the RGB phase), and preventing the object evolving into the HB or AGB phases. Conversely, removing the same fraction of envelope mass on the HB or AGB has almost no effect on the star's subsequent evolution and, in particular, does not noticeably reduce the brightness of the tip of the AGB. The effects of mass loss on main--sequence stars will be discussed in a companion paper and here we concentrate on mass loss due to collisions on the red giant branch.\\
\indent Stellar evolution on the RGB is very fast and stellar properties, particularly core mass and radius (and hence luminosity) change significantly on the RGB. The age of a star is not a useful coordinate against which to measure its evolution and instead, we define a quantity based on the evolution of the core mass, which we use as a proxy for time:
\begin{eqnarray}
\tau_{\mathrm{core}}(t)=\frac{M_{\mathrm{core}}(t)-M_{\mathrm{core}}({\rm BRGB})}{M_{\mathrm{core}}({\rm TRGB})-M_{\mathrm{core}}({\rm BRGB})}
\label{eqn:clobber_coordinate}
\end{eqnarray}
where $M_{\mathrm{core}}({\rm BRGB})$ is the core mass at the base of the red giant branch and $M_{\mathrm{core}}({\rm TRGB})$ is the core mass at the tip. To illustrate the relationship between $\tau_{\mathrm{core}}$ and the actual age of the giant, we plot $\tau_{\mathrm{core}}$ as a function of time \textit{on the RGB} for a $1$\,M$_{\odot}$ giant in Figure \ref{fig:clobberco_realtime}. We show in Figure \ref{fig:giant_1msun_clobber} the lifetimes of $1$\,M$_{\odot}$ giant stars in the middle band ($12>K>10.5$) after encounters at five different positions along the giant branch (characterised by $\tau_{\mathrm{core}}$ on the $y$--axis) which leave them with between $1$ and $100\%$ of their envelopes remaining.\\
\indent The mass loss required to prevent the $1$\,M$_{\odot}$ giant evolving to become visible in the middle band increases as the object ascends the giant branch. As the core mass grows, its later evolution becomes less and less sensitive to the envelope mass. However, even near the base of the giant branch, the quantities of mass that must be removed to seriously affect the giant's evolution are substantial -- several tens of percent at least.\\
\section{Collisions involving giant stars}
\subsection{Collisions of giants with stellar--mass black holes}
\begin{figure}
\includegraphics[width=0.45\textwidth]{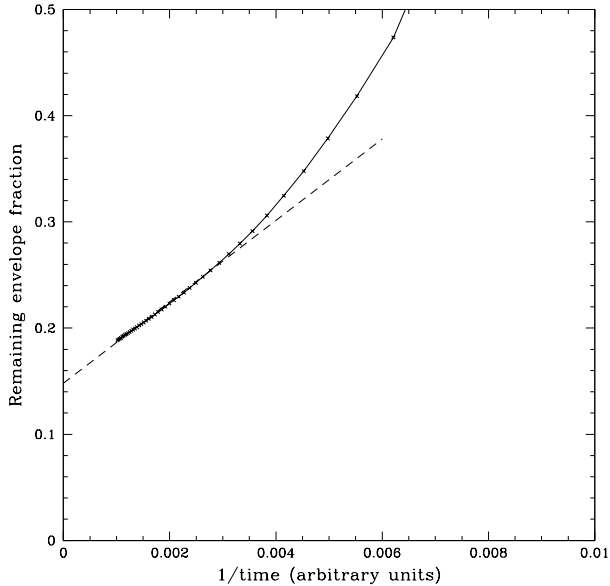}
\caption{Plot of the envelope fraction remaining against $1/$time for the an encounter between a $1$\,M$_{\odot}$ giant and a $10$\,M$_{\odot}$ black hole with a $v_{\infty}$ of $800$\,km\,s$^{-1}$ at an $R_{\rm min}$ of $10$\,R$_{\odot}$. The solid line with the crosses is the measured envelope fraction remaining from the SPH calculation and the dashed line is the fit applied to estimate the final remaining envelope fraction.}
\label{fig:one_over_t_fit}
\end{figure}

\begin{figure}
\includegraphics[width=0.43\textwidth]{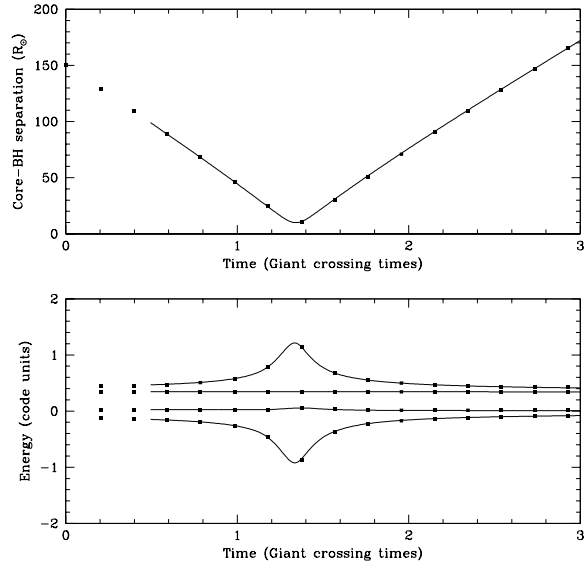}
\caption{Plot comparing the evolution of the separation between the red giant core and the black hole (top panel) and the energies of the system (bottom panel, from top to bottom, kinetic, thermal, total ad gravitational potential energies) of high--resolution and low--resolution simulations of a collision between a $1$\,M$_{\odot}$ giant star and a $10$\,M$_{\odot}$ black hole at a periastron of $10$\,R$_{\odot}$ and $v_{\infty}$ of $400$\,km\,s$^{-1}$. Solid lines are results from the high--resolution calculation and black squares are from the low--resolution calculation. Note that the initial separation between the giant and the black hole in the high--resolution calculation was smaller to save computer time.}
\label{fig:res_compare}
\end{figure}

\begin{figure*}
     \centering
     \subfigure[Low--resolution run ($1.75\times10^{5}$ particles)]{
          \label{fig:low_res}
          \includegraphics[width=.45\textwidth]{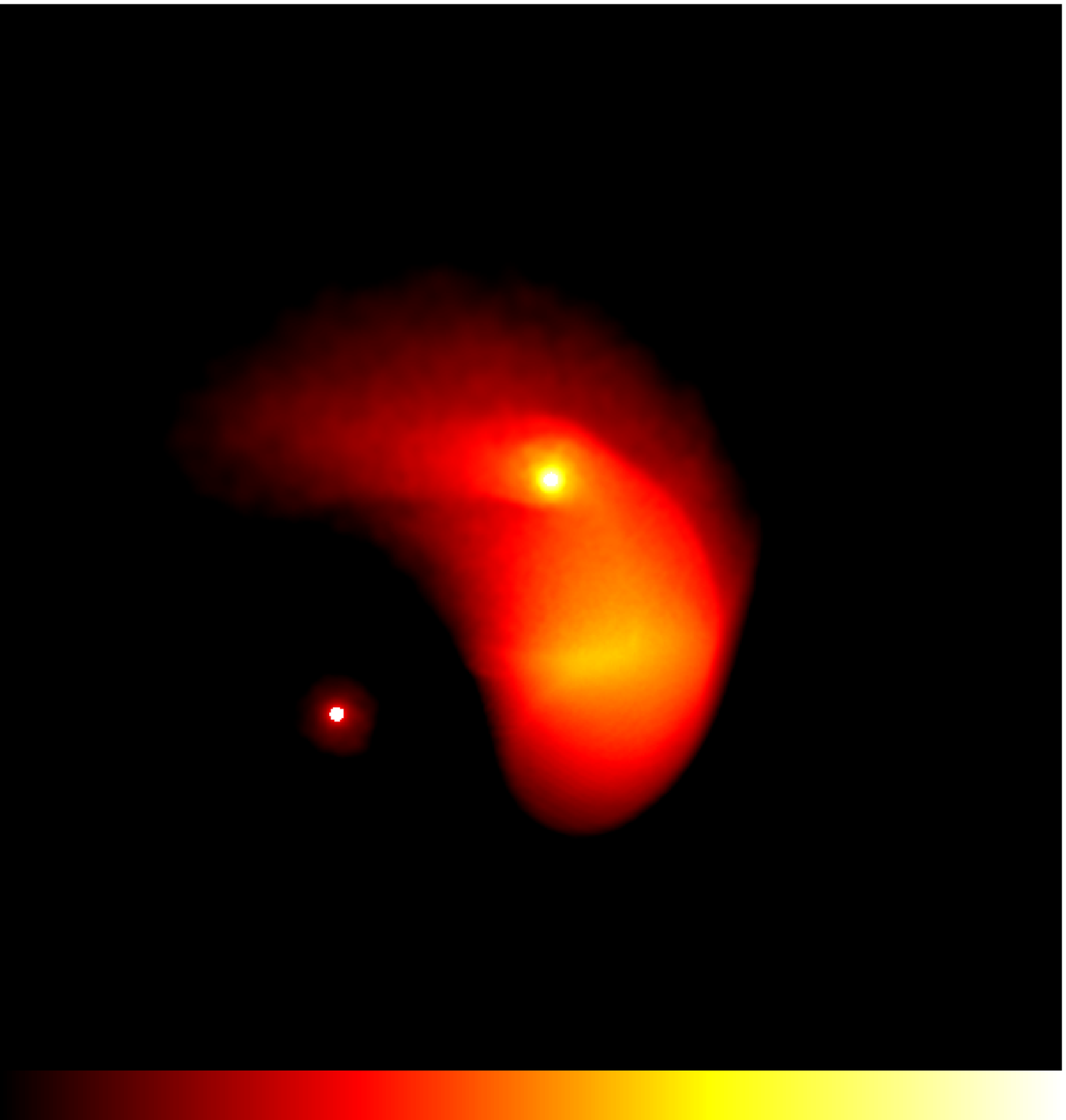}}
     \hspace{.3in}
     \subfigure[High--resolution run ($1.19\times10^{6}$ particles)]{
          \label{fig:high_res}
          \includegraphics[width=.45\textwidth]{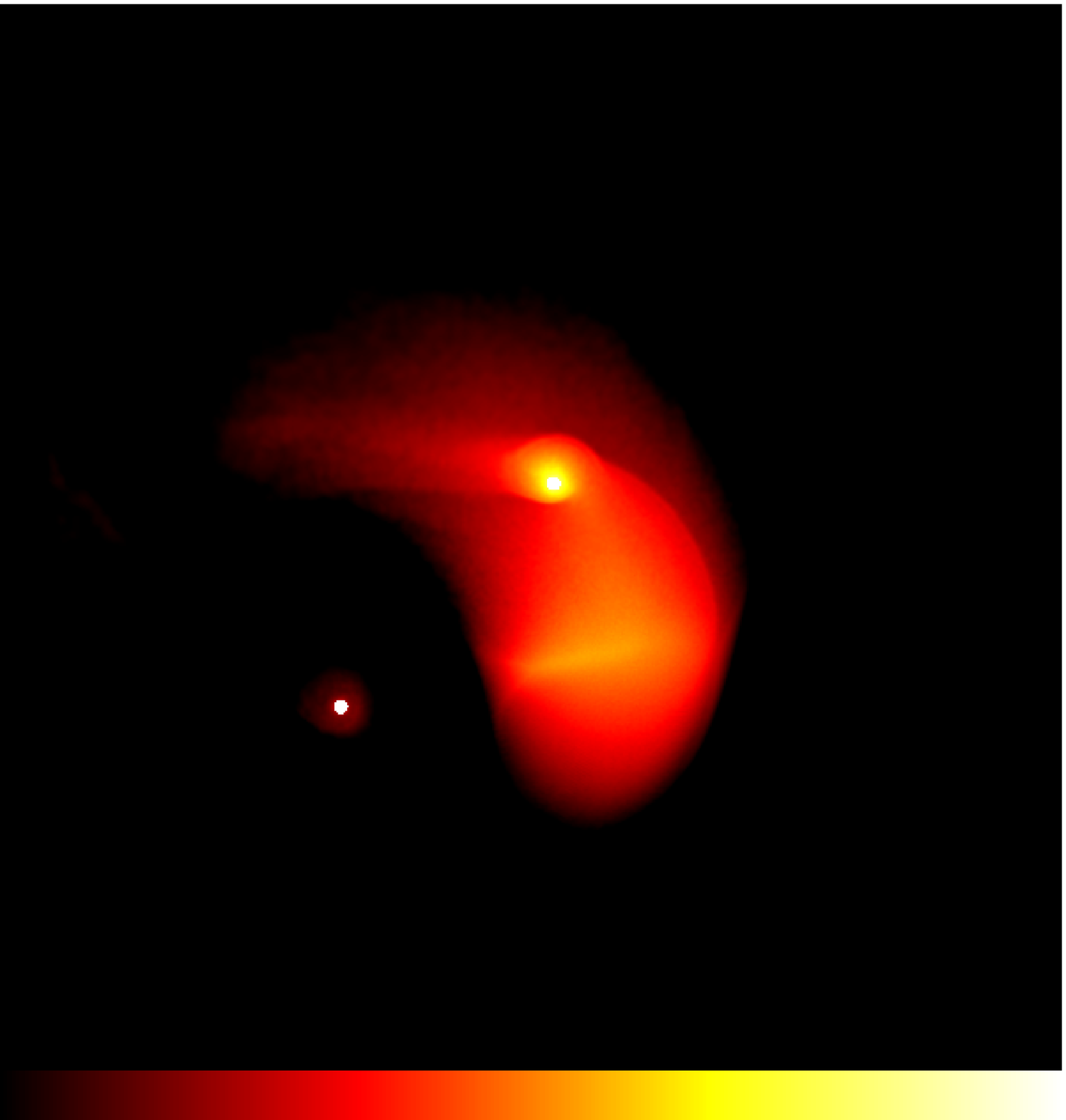}}\\     
     \caption{Comparison of density slices taken through the $z=0$ plane of snapshots from the low-- and high--resolution runs 1.6 crossing times after periastron passage in the encounter between a $1$\,M$_{\odot}$ giant star and a $10$\,M$_{\odot}$ black hole at a periastron of $10$\,R$_{\odot}$ and $v_{\infty}$ of $400$\,km\,s$^{-1}$. Yellow colours represent densities of $6\times10^{-2}$\,g\,$cm^{-3}$ and red represents densities of $6\times10^{-5}$\,g\,cm$^{-3}$. The black hole (moving from right to left) and and the giant core are represented by white dots. Images are $150$R$_{\odot}$ on a side. The ejection of the core and retention of some of the envelope is clearly visible. In this calculation, the envelope fraction retained by the core was $\sim15\%$ in the low--resolution run and $\sim17\%$ in the high--resolution run, which is sufficiently small that the giant will not evolve to become visible in the middle band.}
     \label{fig:screenshot_compare}
\end{figure*}

\begin{figure}
\includegraphics[width=0.45\textwidth]{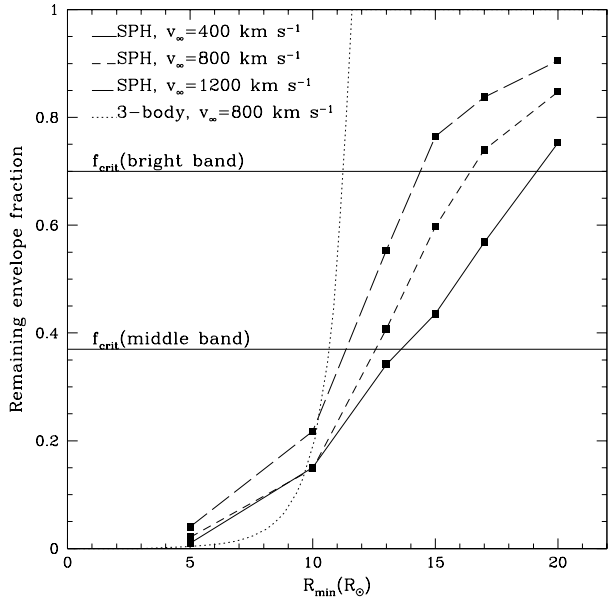}
\caption{Plot of the results of collisions between a $1$\,M$_{\odot}$ giant star and a $10$\,M$_{\odot}$ black hole (black squares) at velocities of $400$ (solid line), $800$ (short--dashed line) and $1200$ (long--dashed line)\,km\,s$^{-1}$. The periastron of the collision is given on the $x$--axis and fraction of the envelope remaining after the collision is given on the $y$--axis. Black lines with squares are the results from SPH calculations. The dotted line is the predicted envelope remaining from the reduced three--body treatment, including the sound speed inside the envelope when calculating the capture radius, taking $v_{\infty}=800$\,km\,s$^{-1}$. The horizontal lines represents the maximum envelope fraction remaining with which this giant does not evolve to become visible in the middle and bright bands at the Galactic Centre.}
\label{fig:mass_loss_1Msun_giant_BH}
\end{figure}

\begin{figure}
\includegraphics[width=0.45\textwidth]{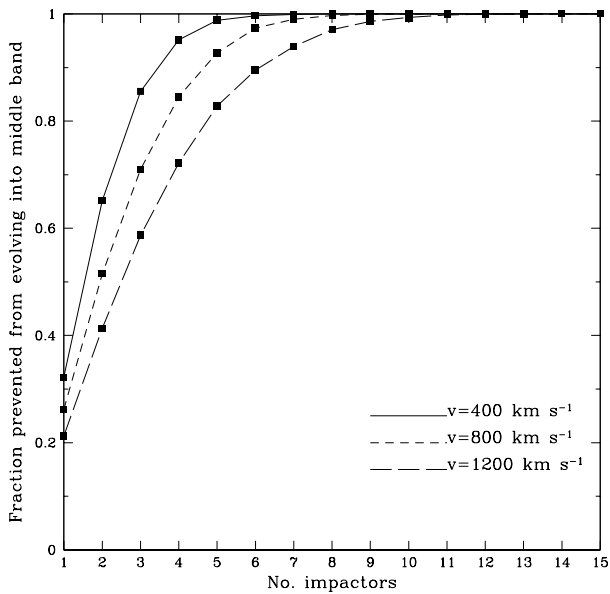}
\caption{Plot of the fraction of $1$\,M$_{\odot}$ giants prevented from evolving to become visible in the middle band against the number of collisions with $10$\,M$_{\odot}$ black holes, assuming that the mass loss as a function of collision periastron is given by the results of SPH simulations at velocities of $400$ (solid line), $800$ (short--dashed line) and $1200$ (long--dashed line)\,km\,s$^{-1}$.}
\label{fig:mc_coll_1Msun}
\end{figure}

\begin{figure}
\includegraphics[width=0.45\textwidth]{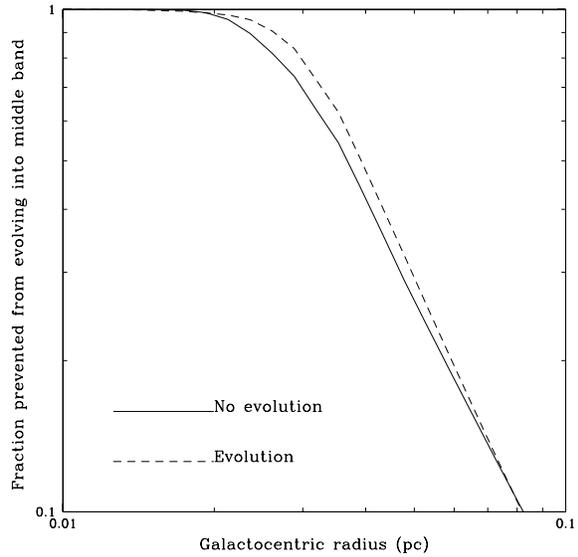}
\caption{Fraction of $1$\,M$_{\odot}$ giants prevented from evolving to become visible in the middle band as a function of Galactocentric radius, assuming that all giants are the `standard' model with $\tau_{\mathrm{core}}=0.51$ (solid line), or assuming giants are randomly distributed in age along the giant branch (dashed line). The vertical dotted line represents the radius within which the middle band is depleted.}
\label{fig:1Msun_evcompare}
\end{figure}

The depleted K--bands reported by \cite{1996ApJ...472..153G} are dominated by $1$, $1.5$, $2$ and $3$\,M$_{\odot}$ stars, as shown in Figure \ref{fig:square_stars}. We chose to model collisions involving  giants approximately halfway up the giant branch, as measured by the quantity $\tau_{\mathrm{core}}$, taking these objects to be representative of giants of a given mass. This has the advantage that such objects are not yet bright enough to be visible in the depleted K--bands and that they lie close to the point at which the cumulative probability of suffering a collision \textit{while on the giant branch} is approximately $0.5$. We give the parameters of our model giants in Table 1.\\
\begin{table*}
\begin{center}
\begin{tabular}{|l|l|l|l|l|l|}
M(M$_{\odot}$) & M$_{\mathrm{core}}($M$_{\odot})$ & R(R$_{\odot}$) & Age(yr) & $\tau_{\mathrm{core}}$ & K(mag)\\
\hline
1.0 & 0.30 & 25.2 & $1.22\times10^{10}$ & 0.51 & 14.5\\
1.5 & 0.29 & 18.8 & $2.95\times10^{9}$ & 0.62 & 15.5\\
2.0 & 0.34 & 31.9 & $1.21\times10^{9}$ & 0.54 & 13.4\\
3.0 & 0.39 & 27.6 & $3.76\times10^{8}$ & 0.40 & 13.6
\end{tabular}
\caption{Properties of giants used in our SPH calculations.}
\end{center}
\label{tab:giant_props}
\end{table*}
\indent The quantities of interest in these simulations are the masses of gas bound to the giant core, and captured by the intruding black hole. To calculate these masses, we employed an iterative scheme. For each point mass, the gas particles in its immediate vicinity were fetched, their velocities relative to the point mass determined and hence their kinetic, potential and thermal energies in the frame of the point mass were calculated. Those particles with negative total energy were taken to be bound to the point mass. A composite object was then constructed from the point mass and those particles bound to it. The mass, centre of mass and centre--of--mass velocity of the composite object were calculated and the procedure repeated until the mass of the composite object converged.\\
\indent After a sharp decrease around the time of the black hole impact, the mass retained by the giant core declined at an ever--decreasing rate for long periods of time. To avoid running simulations for prohibitively long times, we devised a procedure to estimate the final mass of the collision product. Plotting the fraction of the envelope bound to the core, $f_{\mathrm{env}}$ as a function of $1/t$ revealed that, except near the time of the impact, $f_{\mathrm{env}}$ could be well fit by a straight line and extrapolated by least squares fitting to $1/t=0$ (i.e. $t=\infty$), yielding the final envelope mass retained by the core. In Figure \ref{fig:one_over_t_fit}, we show an example plot, taken from the collision of a $1$M$_{\odot}$ with a $10$M$_{\odot}$ black hole at an $R_{\rm min}$ of $10$R$_{\odot}$ and a $v_{\infty}$ of $800$km s$^{-1}$. This plot shows that at the late stages of the collision (i.e. for small $1/t$), the curve is very well fit by a straight line, giving a robust estimate of the final mass loss from the encounter.\\
\indent To check for numerical convergence, we repeated several calculations at much higher resolution on the UKAFF facility at the University of Leicester. In Figure \ref{fig:res_compare} we compare the evolution of the orbital separation (top panel) of the giant core and impacting black hole, and of the energies of the system (lower panel), in the low--resolution calculation (black squares) and the high--resolution calculation (solid lines) of an encounter between a $1$\,M$_{\odot}$ giant and a $10$\,M$_{\odot}$ black hole with a $v_{\infty}$ of $400$\,km\,s$^{-1}$ at an $R_{\rm min}$ of $10$\,R$_{\odot}$. The agreement between the two calculations on the gross evolution of the system is clearly excellent.\\
\indent In Figure \ref{fig:screenshot_compare}, we show density slices in the $z=0$ plane $1.6$ crossing times after periastron passage. The agreement between the low--resolution (left panel) and high--resolution (right panel) resolution calculations is again very good. The same structures are visible in both images, including the disk of captured material around the black hole, and the wake of the black hole through the red giant, including the hole where it has exited the envelope. We also find that the low-- and high--resolution runs agree well on the fraction of the envelope retained by the core, giving $15\%$ and $17\%$ respectively in these calculations.\\
\indent We observed, particularly at smaller periastrons, that the mass loss was often due to the core being ejected and carrying part of the envelope away with it. This is clearly visible in Figure \ref{fig:screenshot_compare}. This phenomenon was pointed out by \cite{1988ApJ...332..271L} and is essentially due to the inability of the envelope to react to the supersonic impact of the intruding black hole. We use our restricted three--body code to determine the velocity of the core relative to the envelope and estimate the mass retained by the mass within the Bondi--Hoyle--Lyttleton radius corresponding to that velocity.\\
\indent For each of our SPH giant--black hole encounters at a given $v_{\infty}$, we performed reduced three--body calculations over a range of periastrons and calculated the mass retained by the core using the Bondi--Hoyle/Bondi--Hoyle--Lyttleton formalism. In Figure \ref{fig:mass_loss_1Msun_giant_BH}, we show the results of SPH calculations of collisions between a $1$\,M$_{\odot}$ giant star and a $10$\,M$_{\odot}$ black hole performed at a range of periastrons and at $v_{\infty}=400$, $800$ and $1200$\,km\,s$^{-1}$ (black lines with squares) compared with the results of our reduced three--body simulations over a similar range of periastrons and with $v_{\infty}=800$\,km\,s$^{-1}$.\\
\indent The SPH simulations and the reduced three--body analysis agree very well at small periastrons ($\lesssim10$\,R$_{\odot}$), but depart at larger values, with the SPH results giving systematically smaller fractions of the envelope being retained (i.e. predicting that more mass is expelled).\\
\indent At the high velocities considered in this paper, there are two regimes of mass loss in interactions between giants and stellar--mass black holes, delineated by the periastron of the collision. At large periastrons $(\gtrsim R_{\rm giant}/2)$, mass loss is largely due to the shock driven by the impactor expelling parts of the giant's outer envelope, leaving the core region relatively undisturbed. Conversely, at periastrons $\lesssim R_{\rm giant}/2$, the core receives a strong impulse and is displaced from the envelope, carrying some of the envelope material away and leaving the rest of the giant to disperse on a dynamical timescale.\\
\indent The critical periastron, inside which collisions expel enough mass to prevent the $1$\,M$_{\odot}$ giant evolving to become brighter than $K=12$ at the Galactic Centre, is predicted to be $\sim12$\,R$_{\odot}$ by the SPH calculations and $\sim10$\,R$_{\odot}$ by the Bondi-Hoyle analysis, resulting in a significant difference in the critical collision cross section if collisions are unfocussed. Hydrodynamic effects thus cannot be neglected in this problem and SPH simulations are essential in determining the collision parameters required to prevent giants evolving into the brighter K--bands in the Galactic Centre.\\
\indent Since little is known about the mass function of black holes, we explored the consequences of impacts with black holes of different mass, to see if, for a black hole population of a fixed \textit{total} mass a smaller number of higher--mass black holes would have a greater affect on a population $1$\,M$_{\odot}$ giants than a larger number of lower--mass black holes. For this to be the case, the critical encounter periastron $R_{\mathrm{critical}}$ for stripping enough of the giant's envelope to prevent it becoming visible in the depleted bands must increase faster than $R_{\mathrm{critical}}\propto\sqrt{M_{\mathrm{BH}}}$, if collisions are not focussed. We found that this is not the case and hence that the increase in $R_{\mathrm{critical}}$ with black hole mass was not enough to offset the smaller number of impactors and therefore that a population of black holes with the canonical mass of $10$\,M$_{\odot}$ is likely to have the greatest effect on the giant population.\\
\indent In collisions with $10$\,M$_{\odot}$ black hole, the critical periastron required to prevent the $1$\,M$_{\odot}$ giant evolving to become brighter than $K=12$ is $\sim R_{\mathrm{giant}}/2$. Hence, if collisions are unfocussed, $\sim25\%$ of single encounters will prevent the giant evolving to become brighter than $K=12$ at the Galactic Centre. This fraction is, however, a lower limit. The collision rates depicted in Figure \ref{fig:coll_rates} imply that giant stars can expect to experience $>1$ encounter with stellar--mass black holes in the innermost $0.1$pc of the Galactic Centre, roughly the region where the population of giants is observed to be depleted. In addition, Figure \ref{fig:mass_loss_1Msun_giant_BH} shows that significant mass loss occurs at periastrons larger than the critical one, so several successive encounters may remove a large enough fraction of the a giant's envelope to affect it to the degree required.\\
\indent We used our Monte Carlo code, with black--hole impactors alone, to generate the fraction of giants prevented from evolving into the middle and bright bands for a given number of impactors at three velocities for the $1$ and $2$\,M$_{\odot}$ giants and a single velocity for the $1.5$ and $3$\,M$_{\odot}$ giants. The results of our calculations for the $1$\,M$_{\odot}$ giant are shown in Figure \ref{fig:mc_coll_1Msun}.\\
\indent A given Galactocentric radius uniquely determines a relative velocity and cumulative impact probability. To map the fractions of objects prevented from evolving into the bright and middle bands, we performed two--dimensional interpolation in $v_{\infty}$ and collision probability for the $1$ and $2$\,M$_{\odot}$ giants to determine the fraction of these objects prevented from evolving into the middle and bright bands as a function of Galactocentric radius. For the $1.5$ and $3$\,M$_{\odot}$ giants, we interpolated in collision probability only.\\
\indent This analysis assumes that all giants of a given mass may be represented by the models at the halfway stage on the giant branch used in the SPH calculations. To check the influence of the evolution of the giants, we repeated the Monte Carlo simulations using $10^{4}$ $1$\,M$_{\odot}$ giants with ages randomly distributed on the giant branch. Once the age of each giant was chosen, a three--body mass--loss/$R_{\mathrm{min}}$ curve was generated by interpolating between the mass--loss curves generated from the three--body calculations using giants bracketing the chosen age. As Figure \ref{fig:mass_loss_1Msun_giant_BH} shows, the three--body calculations underestimate the mass loss at large values of $R_{\mathrm{min}}$ where hydrodynamic effects become important. We allowed for this by calculating from Figure \ref{fig:mass_loss_1Msun_giant_BH} the factor by which the $R_{\mathrm{min}}$ resulting in a given fraction of the giant envelope was larger in the SPH results than in the three--body results.  Assuming that these factors were the same for giants of any age, we produced a synthetic mass--loss curve for each giant of the randomly--chosen age. Collisions with these giants and a range of numbers of black hole impactors were then performed in the same way and the fraction of giants prevented from evolving into the middle and bright bands as a function of Galactocentric radius determined. In Figure \ref{fig:1Msun_evcompare}, we compare the fraction of $1$\,M$_{\odot}$ giants prevented from evolving into the middle band as a function of Galactocentric radius, taking all giants to be the representative $\tau_{\mathrm{core}}=0.51$ model (solid line), or accounting for the giant's evolution in the way described. We see that the giant's evolution has little influence on the results and that the models we chose to perform hydrodynamic simulations with are well representative of giants of each given mass.\\
\indent We see that collisions with black holes alone reduce the population of giants in the middle band by $\gtrsim50\%$ inside a radius of $0.04$ pc, comparable to the size of the region depleted in this band. However, we will not draw any conclusions about the overall effect of collisions on the giant population until we have considered collisions with main sequence stars, described in the next section.\\
\indent In all of our SPH calculations, some material was captured by the black hole during its passage though the giant's envelope, the captured gas remaining in orbit in a disc--like structure around the hole. The black hole will accrete this material and the resulting energy release will make it visible as an X--ray source. In order to see how many such sources we might expect to observe, we assume that the captured material will be accreted at the Eddington rate, given by $\dot{M}=2\times10^{7}$\,M$_{\odot}$\,yr$^{-1}$ for a $10$\,M$_{\odot}$ black hole. The black holes typically capture a few$\times10^{-2}-10^{-1}$\,M$_{\odot}$ of material, so the time for which they will be visible as accreting X--ray sources is $\sim5\times10^{4}-5\times10^{5}$\,yr. If we assume that the collision rate per giant is given simply by $n_{\mathrm{BH}}\sigma_{\mathrm{giant}}v_{\infty}$ and take fiducial values appropriate for the innermost $0.1$\,pc of $n_{\mathrm{BH}}=10^{6}$\,pc$^{-3}$, $\sigma_{\mathrm{giant}}=\pi R_{\mathrm{giant}}^{2}$ (neglecting gravitational focussing) with $R_{\mathrm{giant}}=50$\,R$_{\odot}$ and $v_{\infty}=1000$\,km\,s$^{-1}$, we obtain a collision rate per giant of $4\times10^{-9}$\,yr$^{-1}$. Our Galactic Centre models imply that the central $0.1$\,pc should contain $\sim10^{4}$ giants, so that the rate of giant--BH collisions within this volume would be $4\times10^{-5}$\,yr$^{-1}$. Since this figure is approximately the inverse of the accretion timescale, we would expect to observe one or zero black holes accreting material which they have acquired during a collision with a giant. This result is consistent with \cite{2005ApJ...622L.113M}.\\
\subsection{Collisions of giants with main--sequence stars}
\indent Encounters between giants and MS stars have already been studied in the context of the giant depletion in the Galactic Centre by \cite{1999MNRAS.308..257B}. However, the two giant stars studied by \cite{1999MNRAS.308..257B} were both in quite extreme evolutionary stages. Their $2$\,M$_{\odot}$ giant was near the tip of the RGB and their $8$\,M$_{\odot}$ giant near the tip of the AGB. We have chosen to study encounters with giants towards the middle of the RGB which are representative of all giants of a given mass, so we extend our study to include encounters between giants and $1$\,M$_{\odot}$ MS impactors, which we again treat as point masses. In addition, we extend the work of \cite{1999MNRAS.308..257B} by considering the cumulative effect of multiple impacts. We do not consider white dwarf or neutron star impactors. This is partly because, from a numerical point of view, there is little difference between a white dwarf, a neutron star and a low--mass main--sequence stars, since they are all point masses of similar mass. In addition, our population synthesis models suggest that that MS stars outnumber white dwarfs by $\sim4:1$ and that neutron stars are even less numerous. A flatter IMF than the Miller--Scalo one we have used could increase the ratio of white dwarfs to MS stars but we found that such an IMF was not well able to reproduce the observed Galactic Centre giant population. We do not think it is likely that the number of white dwarfs in the Galactic Centre exceeds the number of MS stars. It is also possible that a significant fraction of neutron stars would be ejected from the Galactic Centre by their natal kicks.\\
\begin{figure}
\includegraphics[width=0.45\textwidth]{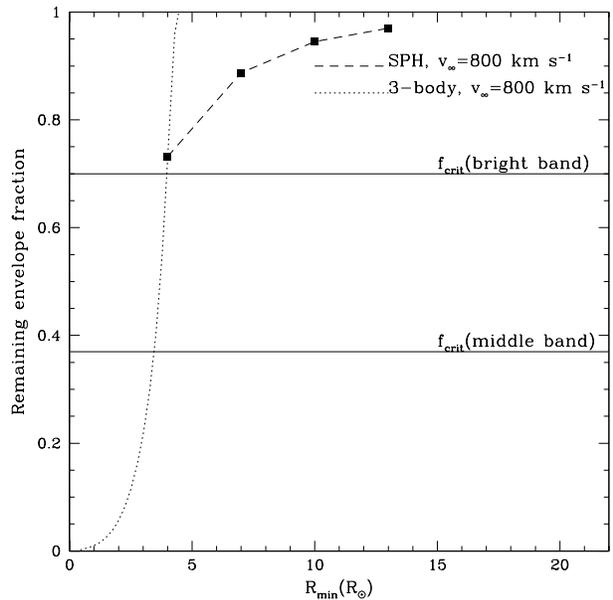}
\caption{Plot of the fractional envelope mass remaining after collisions between a $1$\,M$_{\odot}$ giant star and a $1$\,M$_{\odot}$ main sequence star (black squares) at a velocity of $800$\,km\,s$^{-1}$. The periastron of the collision is given on the $x$--axis and fraction of the envelope remaining after the collision is given on the $y$--axis. The dotted line is the predicted envelope remaining from the reduced three--body treatment, including the sound speed inside the envelope when calculating the capture radius. The horizontal lines represents the maximum envelope fraction remaining with which this giant does not evolve to become visible in the middle and bright bands at the Galactic Centre.}
\label{fig:1MsunRGB_1MsunMS}
\end{figure}
\indent The smallest values of $R_{\mathrm{min}}$ considered by \cite{1999MNRAS.308..257B} corresponded to $\approx R_{\mathrm{giant}}/4$ for their two models. We repeated our three--body/Bondi--Hoyle analysis using $1$\,M$_{\odot}$ impactors and found that collisions in which $R_{\mathrm{min}}<R_{\mathrm{giant}}/6$ may eject the giants' cores and therefore strip the cores of sufficient mass to affect the giants' evolution in a similar manner to encounters with black holes at larger periastrons. Although the critical encounter cross section for a MS impactor to have similar effects to a BH impactor is then much smaller, the number density of MS stars at the Galactic Centre is much larger than that of black holes, so that encounters with MS stars may have an effect on the giant population comparable to or greater than those with black holes.\\
\begin{figure}
\includegraphics[width=0.45\textwidth]{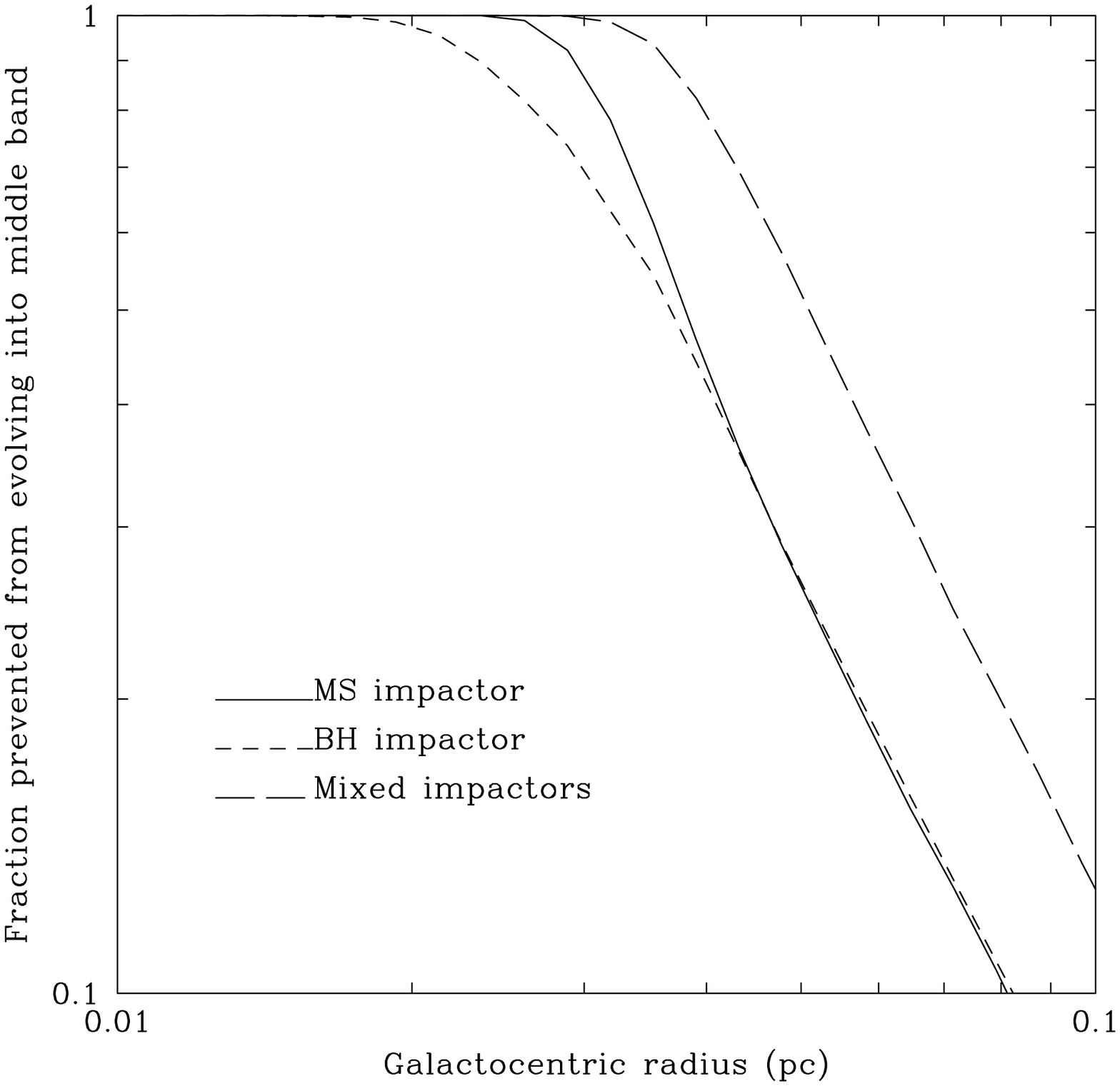}
\caption{Fraction of $1$\,M$_{\odot}$ giants prevented from evolving into the middle band as a function of Galactocentric radius by collisions with MS stars (solid line) and BHs (short--dashed line), assuming collision rates boosted by a factor of two due to eccentricity of giant's orbit around GC, and using 2--D interpolation to determine number of impactors, relative velocity and thus kill fraction as functions of Galactocentric radius for the BH impactors and assuming a single velocity of $800$km s$^{-1}$ for the MS impactors. The long--dashed line is the fraction of $1$\,M$_{\odot}$ giants prevented from evolving into middle band by a mixed population of black holes and main sequence stars.}
\label{fig:1MsunMSBH_mid_clobber}
\end{figure}
\begin{figure}
\includegraphics[width=0.45\textwidth]{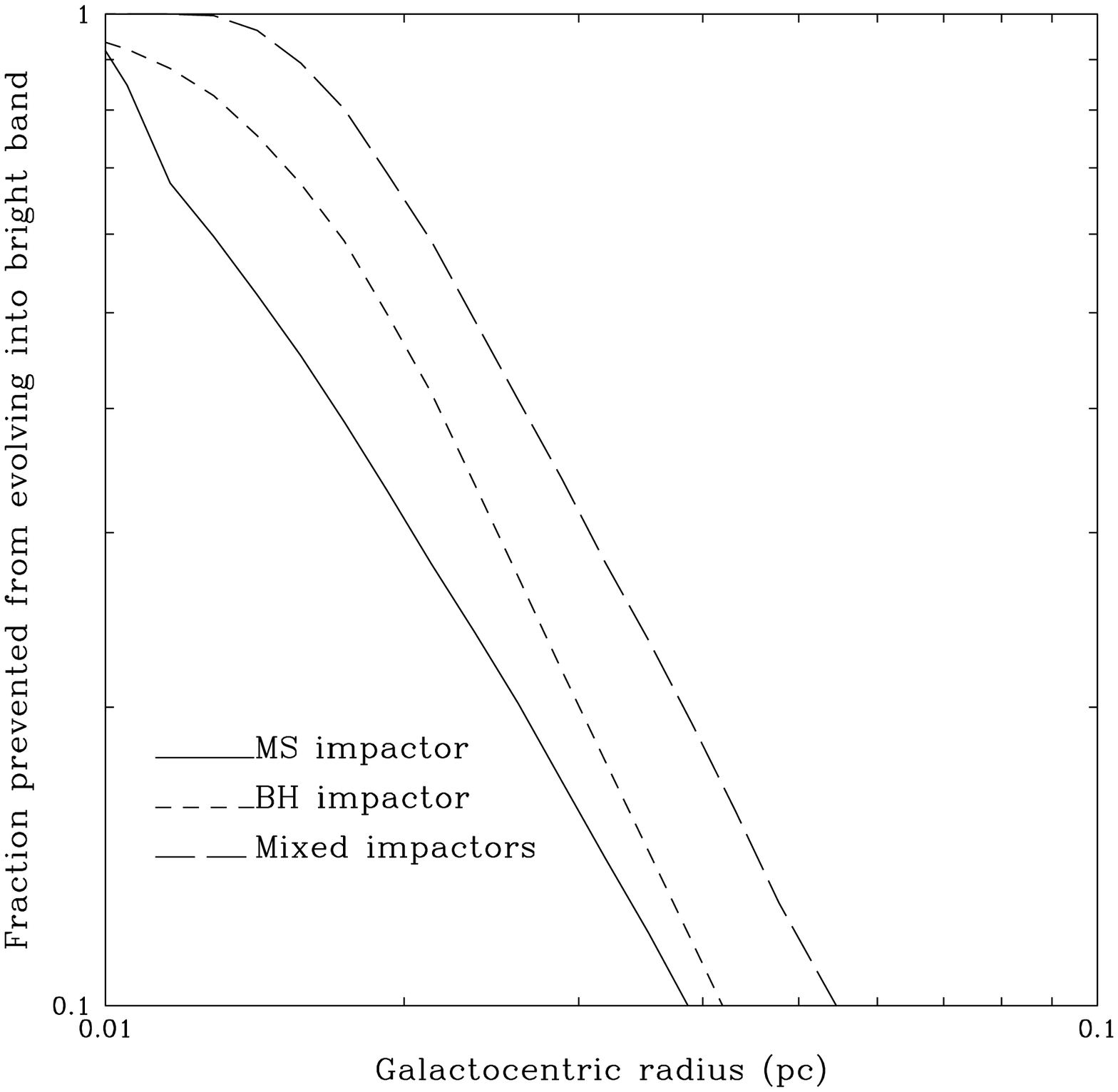}
\caption{Fraction of $2$\,M$_{\odot}$ giants prevented from evolving into the bright band as a function of Galactocentric radius by collisions with MS stars (solid line) and BHs (short--dashed line), assuming collision rates boosted by a factor of two due to eccentricity of giant's orbit around GC, and using 2--D interpolation to determine number of impactors, relative velocity and thus kill fraction as functions of Galactocentric radius for the BH impactors and assuming a single velocity of $800$\,km\,s$^{-1}$ for the MS impactors. The long--dashed line is the fraction of $2$\,M$_{\odot}$ giants prevented from evolving into the bright band by a mixed population of black holes and main sequence stars.}
\label{fig:2MsunMSBH_brt_clobber}
\end{figure}
\indent We performed SPH simulations to check the results of our restricted three--body analysis. In agreement with \cite{1999MNRAS.308..257B}, we found that encounters at $R_{\mathrm{min}}>R_{\mathrm{giant}}/4$ eject little mass, at most $\sim10\%$ of the envelope. However, as shown in Figure \ref{fig:1MsunRGB_1MsunMS}, the SPH results and the three--body results both suggest that single impacts at smaller periastrons may be able to expel significant quantities of mass, or eject the giant core. We caution, however, that we have not performed SPH simulations of encounters between giants and MS stars in which the separation between the giant core and the centre of mass of the MS was less than $4$\,M$_{\odot}$ since the treatment of the MS star as a point mass may lead to misleading results at such small separations.\\
\indent In addition, Figure \ref{fig:coll_rates} suggests that giants can expect to suffer several tens to of order $100$ impacts with MS stars at Galactocentric radii $<0.1$\,pc. We therefore repeated the Monte Carlo calculations from the previous section allowing for multiple impacts with a mixture of MS and BH impactors. We used the collision probabilities from Figure \ref{fig:coll_rates} to determine the expected numbers of black hole and main sequence impactors at a given Galactocentric radius. We then performed Monte Carlo calculations and calculated the cumulative effect of all the impactors. In Figure \ref{fig:1MsunMSBH_mid_clobber} we show the results of this analysis for the $1$\,M$_{\odot}$ giant. The figure depicts, as a function of Galactocentric radius, the fraction of $1$\,M$_{\odot}$ giants prevented from evolving to become visible in the middle band, assuming BH impactors acting alone (short--dashed line), assuming MS impactors acting alone (solid line) and assuming a mixture of BH and MS impacts in proportions dictated by the relevant collision probabilities (long--dashed line). Figure \ref{fig:2MsunMSBH_brt_clobber} depicts the fraction of $2$\,M$_{\odot}$ stars prevented from evolving into the bright band.\\
\indent We find that the MS stars are approximately as effective -- somewhat more so in the case of the $1$\,M$_{\odot}$ giant -- in changing the visible giant population as the BH impactors. This result demands some explanation.\\
\begin{figure}
\includegraphics[width=0.45\textwidth]{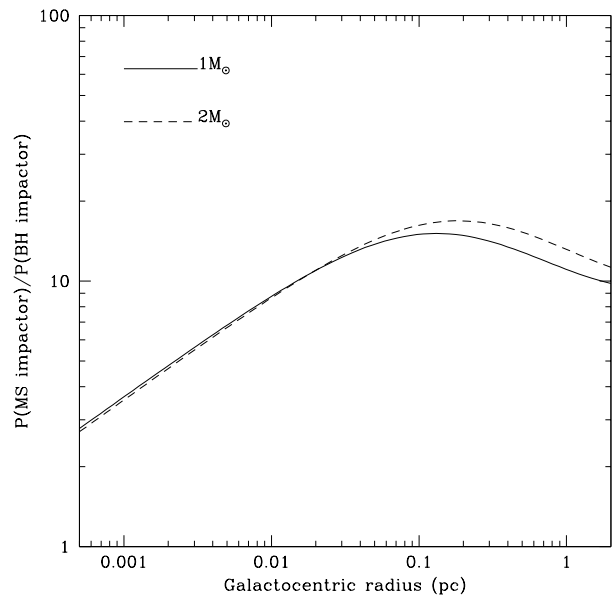}
\caption{Comparison of the ratio of the probability of being struck by a main sequence star to that of being struck by a black hole as a function of Galactocentric radius for a $1$\,M$_{\odot}$ giant (solid line) and a $2$\,M$_{\odot}$ giant (dashed line).}
\label{fig:pcol_ratio}
\end{figure}
\indent In Figure \ref{fig:pcol_ratio}, we plot for $1$ and $2$\,M$_{\odot}$ giants, the ratio as a function of Galactocentric radius, of the probability of the giant being hit by a main sequence star to the probability of it being hit by a black hole. We see that, in the range $0.01$\,pc$>r>0.1$\,pc, this ratio is $\approx10$ for both giants. In Figure \ref{fig:1MsunMSBH_nimp}, we plot the fraction of $1$\,M$_{\odot}$ stars prevented from evolving to become visible in the middle band as a function of the number of impactors, assuming BH impactors alone (dashed line) and MS impactors alone (solid line). We see that the factor by which the number of MS impactors must exceed the number of BH impactors in order to deplete the same fraction of giants is at most $\sim5$. It is therefore not surprising, given the assumptions inherent on our model, that MS impactors are as effective in depleting giants as black holes.\\
\indent The cumulative effect of collisions with black holes and main sequence stars on the $1$\,M$_{\odot}$ giants which dominate the middle band is to evacuate a three--dimensional volume $\sim0.04$\,pc in radius of giants. The effect of collisions on the $2$\,M$_{\odot}$ giants dominating the bright band is smaller, evacuating a region $\sim0.02$\,pc in radius.\\
\begin{figure}
\includegraphics[width=0.45\textwidth]{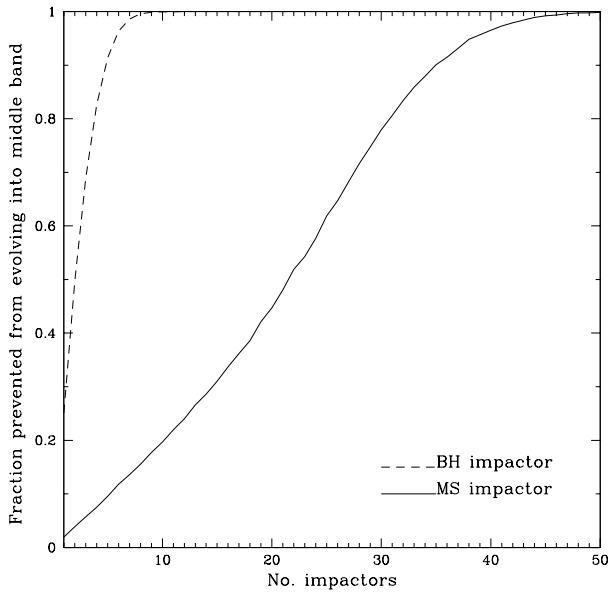}
\caption{Comparison of the fraction of $1$\,M$_{\odot}$ giants prevented from evolving into the middle band against the number or impactors, where each impactor is either a $10$\,M$_{\odot}$ black hole or a $1$\,M$_{\odot}$ main sequence star.}
\label{fig:1MsunMSBH_nimp}
\end{figure}
\section{Discussion}
\subsection{Main sequence versus black hole impactors}
We used the results of Monte Carlo modelling combined with SPH calculations and three--body simulations to quantify the mass lost by a given giant in an encounter at a given $R_{\rm min}$ with a black hole or a main--sequence star. Taking successive collisions to be independent and assuming that BH impactors and MS impactors acted alone or in concert, we found that the effects on the visible giant population of the two types of impactor were comparable, owing to the much greater number density of MS stars than black holes in the Galactic Centre.\\
\indent This result should be treated with some caution, however. In the three--body and SPH calculations, we have treated the impactors as point masses. While this is clearly valid for the black holes, it may not always be so for the MS stars, particularly in the encounters at very small $R_{\rm min}$ required to expel the giant core.\\
\indent The periastron necessary to notionally eject the core may be so small that it is not valid to approximate the MS star as a point mass, either because it is physically too small to fit, or because it approaches the core so closely that it experiences Roche--lobe overflow onto the core.\\
\indent The size of the Roche lobe is given by \citep{1983ApJ...268..368E}
\begin{eqnarray}
r_{R}^{i}=\frac{0.49(m_{i}/m_{j})^{\frac{2}{3}}a}{0.6(m_{i}/m_{j})^{\frac{2}{3}}+\mathrm{ln}(1+(m_{i}/m_{j})^{\frac{1}{3}})}.
\end{eqnarray}
Given the core masses given in Table 1, we find that, for all the giants, the MS star cannot approach the core closer than $2.0-2.2$\,R$_{\odot}$ without experiencing Roche lobe overflow. The consequences of this phenomenon are not obvious. The high--velocity of the intruding MS star may render the amount of mass actually transferred in the flyby with the giant core very small. In this case, the MS star will exit the giant envelope relatively unscathed and the giant core will probably be ejected. However, it is possible that the tidal distortion of the MS star (or simply an impact at a periastron less than the radius of the MS star) will lead to a collision between the core and the MS star. This encounter may dissipate enough energy that the MS star becomes bound to the core and sinks or spirals into the centre of the giant to smother the core. The ultimate result of such an encounter may be that the MS star becomes part of the giant envelope, resulting in a larger and brighter giant. Understanding which (if either) of these scenarios is correct and what effect they would have on the giant's evolution require detailed high--resolution hydrodynamic calculations which are unfortunately beyond the scope of this paper, but would make an interesting topic of further work. Our three--body analysis shows that it becomes significant for giants of $3$\,M$_{\odot}$ and above, as the $R_{\rm min}$ required for a $1$\,M$_{\odot}$ object to expel core is $\approx2$\,R$_{\odot}$. The fractions of giants prevented from evolving to become visible in the depleted K--bands by MS impactors should therefore be regarded as upper limits and it may be that encounters with black holes are in reality significantly more important.\\
\subsection{Encounters with main--sequence binaries}
\indent \cite{1998MNRAS.301..745D} studied the effects of encounters between MS \textit{binaries} and giants. Since binaries have much larger cross sections than single stars, even a relatively small population of binaries could be involved in a significant number of collisions. The maximum binary cross section that can be considered is set by the requirement that the binary must be hard (so that it is not disrupted by encounters) and is therefore determined by the local velocity dispersion. As the velocity dispersion increases, the hard--soft boundary semimajor axis and therefore the cross section of the binary decreases. Additionally, \cite{1998MNRAS.301..745D} found that, at higher relative velocities, a smaller fraction of interactions resulted in outcomes destructive to the giant. The relative velocities considered by \cite{1998MNRAS.301..745D} ranged from $50-150$\,km\,s$^{-1}$. We extrapolate their results to the velocities considered in this paper of $\sim1000$\,km\,s$^{-1}$. Following \cite{1983ApJ...268..319H} and \cite{1998MNRAS.301..745D}, we define the maximum interaction separation $s_{\rm max}$ as a function of $v_{\infty}$ as 
\begin{eqnarray}
s_{\rm max}=\left(\frac{4.0}{v_{\infty}/v_{\rm crit}}+0.6\right) d
\end{eqnarray}
where $v_{\rm crit}$ is the critical velocity of the chosen binary, defined as the velocity at infinity an intruder of the same mass as the binary components must have such that the total energy of the three--body systems is zero, and $d$ is the binary separation. If we adopt the same binary parameters as \cite{1998MNRAS.301..745D}, $d=23.5$\,R$_{\odot}$ and $v_{\rm crit}=116.7$\,km\,s$^{-1}$. If $v_{\infty}$ is $1000$\,km\,s$^{-1}$, we obtain the cross section of the binary $s_{max}^{2}\sim630$\,R$_{\odot}^{2}$. Even if we make the assumption that \textit{all} interactions at $R_{\rm min}<s_{\rm max}$ prevent the giant evolving into the depleted K--bands, for the $93$\,R$_{\odot}$ $2$\,M$_{\odot}$ giant considered by \cite{1998MNRAS.301..745D}, this is $\approx R_{\rm giant}^{2}/13$. Given that the cross section for a single encounter with a $1$\,M$_{\odot}$ impactor to prevent this giant evolving into the depleted K--bands is $\approx R_{\rm giant}^{2}/36$, this implies that binary interactions are a factor of $\sim3$ more effective in altering the giant population, implying that the stellar population must consist of at least $25\%$ binaries in order that binary encounters are more significant than encounters with single MS stars. In agreement with \cite{1998MNRAS.301..745D}, we find that the fraction of binaries would have to be unrealistically high in the Galactic Centre to have a significant collisional effect on the population of giants there.\\
\subsection{Effect of collisions on the observed giant population}
\begin{figure}
\includegraphics[width=0.45\textwidth]{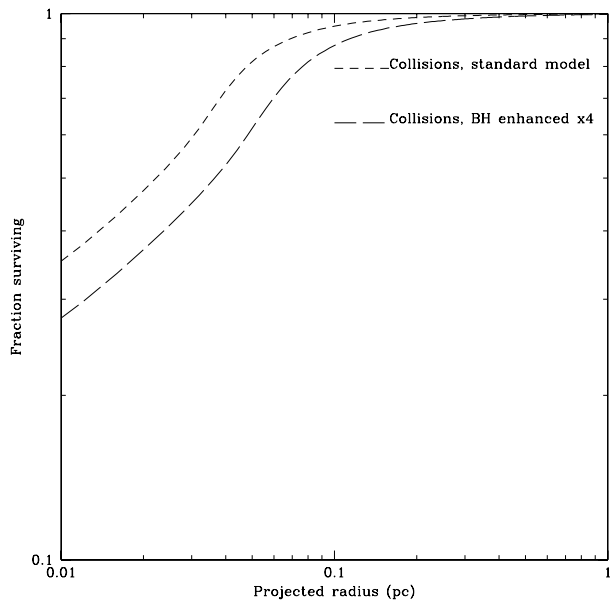}
\caption{Surviving fraction of $1$M$_{\odot}$ giants visible in the middle band plotted against projected Galactocentric radius, assuming our standard BH cusp (short--dashed line) and a cusp with four times as many BH (long--dashed line). We took $r_{\mathrm{max}}$ to be $10$\,pc. Results with $r_{max}=5$\,pc and $r_{\mathrm{max}}=20$\,pc are very similar.}
\label{fig:fclob_proj}
\end{figure}

\indent For a single impact with a given giant, more mass has to be ejected to prevent the giant evolving into the middle band than to prevent it evolving into the bright band, since the former entails a more severe perturbation to the star's evolution. The collision cross section to prevent a giant evolving into the bright band is correspondingly larger than that to prevent the same giant from evolving into the middle band. However, ejecting a given fraction of the envelope from a $2$\,M$_{\odot}$ giant is intrinsically more difficult than ejecting the same envelope fraction from a $1$\,M$_{\odot}$ star, since the $2$\,M$_{\odot}$ giant's envelope is more strongly bound. In addition, the times which stars spend on the giant branch decrease strongly as the stellar mass increases. These two facts largely explain why we find that the bright band is depleted to a similar radius as the middle band. The bright band is dominated by $2-3$\,M$_{\odot}$ giants which have more strongly bound envelopes and which are less likely to be struck by MS or BH impactors while on the giant branch.\\
\indent We have calculated for each model giant the probability at a given Galactocentric radius that a combination of collisions with black holes and MS stars will prevent the giant evolving to become visible in the middle or bright bands. However, in order to compare our results with observations, we must examine the effect of these collisions on the projected surface density of sources at the Galactic Centre, since this is what is actually observed. If we assume that the giant stars are distributed in the same way as the MS stars, i.e. that $n_{\mathrm{giant}}(r)\propto r^{-1.4}$ and that the distribution has a maximum radius $r_{\mathrm{max}}$ sufficiently large that changing it by factors of two does not influence the result significantly, we can simply integrate the surface density $\Sigma$ of stars at any projected radius $x$ as
\begin{eqnarray}
\Sigma(x)=\int_{-z_{\mathrm{max}}}^{+z_{\mathrm{max}}}\left[1-f(z)\right]n(z)dz,
\end{eqnarray}
where $z=\sqrt{(r^{2}-x^{2})}$, $z_{\mathrm{max}}=\sqrt{(r_{\mathrm{max}}^{2}-x^{2})}$ and $f(z)=f(r)$ (where $r=\sqrt{(z^{2}+x^{2})}$) is the fraction of giants prevented from evolving into the relevant band at three--dimensional radius $r$. The quantity of interest is then the ratio, as a function of projected radius $x$, of $\Sigma$(collisions)/$\Sigma$(no collisions), since this reveals the effect of collisions on the observed surface density of sources. The results of performing this analysis on the $1$\,M$_{\odot}$ giant in the middle band are shown in Figure \ref{fig:fclob_proj}. The two curves show the depletion expected due to collisions with MS stars and (i) our standard population of BH (5000 within $0.1$\,pc, short--dashed line), (ii) a black hole population enhanced by a factor of four. We find that the standard population of black holes (in combination with the MS stars) depletes $50\%$ or more giants from the middle band within a \textit{projected} radius of $\sim0.02$\,pc, while the enhanced BH population depletes $50\%$ or more middle band giants within $\sim0.04$\,pc. This is comparable in size to the depleted region observed in the middle band by \cite{1996ApJ...472..153G}. The black hole population required to achieve this is somewhat larger than that predicted by \cite{2004ApJ...606L..21A} and \cite{2007MNRAS.377..897D}.\\
\subsection{Monte Carlo simulations of the Galactic Centre}
\indent To examine the influence of collisions on the Galactic Centre environment in more detail, we used our Monte Carlo population synthesis model of the Galactic Centre cluster (from which Figure \ref{fig:square_stars} was constructed). We assume that star formation there has been proceeding at a constant rate for $14$\,Gyr and that stars are born with a Miller--Scalo IMF. As discussed in Section 2, when we varied these assumptions, we obtained a stellar population that was a significantly worse fit to the observed numbers of giants in all three K--bands throughout the whole Galactic Centre cluster, so we conclude that these assumptions are reasonable. We use the same STARS evolution tracks and \cite{1966ARA&A...4..193J} colours and bolometric corrections to calculate the evolutionary phase and K--magnitude of each star. The stars are distributed in a sphere $2$\,pc in radius according to the power laws derived by \cite{2003ApJ...594..812G} using the normalisation described in Equation \ref{eqn:nms_r}. We constructed model clusters neglecting the effects of stellar collisions, models including collisions with MS stars and our standard and enhanced populations of black holes. We then compared all our models with new observational data obtained from \cite{priv_trippe_2007}.\\
\indent We constructed plots of the mean cumulative number of stars in our Monte Carlo models in the middle band against projected Galactocentric radius and compared them to the same plot generated from the observational data.\\
\begin{figure}
\includegraphics[width=0.45\textwidth]{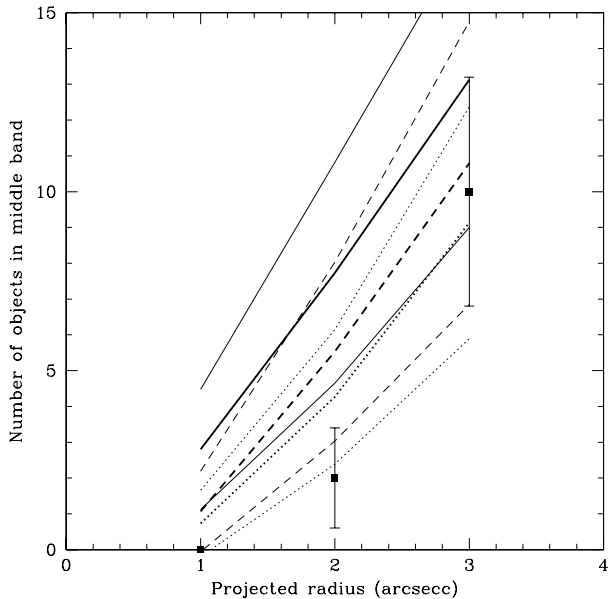}
\caption{Cumulative number of stars against projected Galactocentric radius measured at $0.04$, $0.08$ and $0.12$\,pc. Squares are observed numbers from \protect \cite{priv_trippe_2007} with $\sqrt{n}$ errors. Lines are results of Monte Carlo realisations with no collisions (solid lines), collisions with our standard black hole and main--sequence populations (long--dashed lines) and collisions with a black hole population enhanced by a factor of four (dotted lines). Thick lines show the the mean numbers of stars averaged over fifteen Monte Carlo realisations and thin lines are +/- one standard deviation.}
\label{fig:MC_3arcsec}
\end{figure}
\begin{figure}
\includegraphics[width=0.45\textwidth]{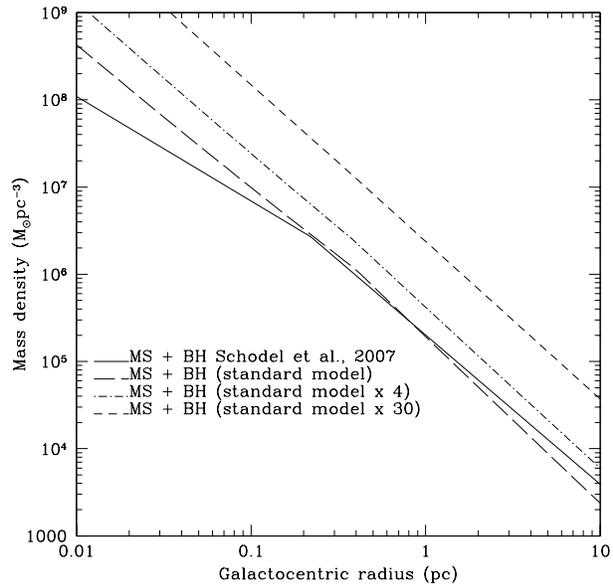}
\caption{Plot of total mass densities against Galactocentric radius for models with our standard BH population (long dashed line), models with black hole populations four (dash--dot line) and thirty (short--dashed line) times larger, and after Sch\"odel et al., (2007), Figure 7 (solid line).}
\label{fig:mass_dens}
\end{figure}
\indent In Figure \ref{fig:MC_3arcsec} we show the cumulative number of stars in the middle band against projected Galactocentic radius generated by our Monte Carlo models (lines with +/- $1\sigma$ standard deviation) and compare these with the observed numbers of objects (squares, also with +/- $1\sigma$ errors). It is clear that both models in which collisions are included (dashed and dotted lines in Figure \ref{fig:MC_3arcsec}) give a much better fit to the observations than a model in which collisions are not included (solid line in Figure \ref{fig:MC_3arcsec}), and that the model with the enhanced black hole population gives a good fit to the observed data. We therefore conclude that collisions with MS stars and black holes may be able to account for the depletion of giants in the middle band, although with a rather large black hole population.\\
\indent Repeating this analysis for the bright band (dominated by $2$\,M$_{\odot}$ giants), we find that the region evacuated of giants by collisions is too small ($\sim0.02$\,pc in radius) to explain the observed depletion, even using our enhanced population of black holes. We determined that the black hole population required to deplete the $2$\,M$_{\odot}$ giants out to a radius of $0.1$\,pc is at least thirty times more populous than our standard model, so that there would be $1.5\times10^{5}$ BH within $0.1$\,pc. The effect of this population would be to deplete the \textit{middle} band giants out to a radius of $\sim0.3$\,pc, which is much greater than the observed depletion radius. This BH population is also much larger than that inferred by, e.g., \cite{2004ApJ...606L..21A}, \cite{2007MNRAS.377..897D}. In addition in Figure \ref{fig:mass_dens}, we plot the total mass density against Galactocentric radius for the model with our standard black hole population and models with black hole populations four and thirty times large, and compare it to that inferred in \cite{2007A&A...469..125S}. We see that, at $0.1$\,pc, the model with four times our standard number of black holes has a mass density $\sim2$ times that inferred by \cite{2007A&A...469..125S} and the model with thirty times the standard number of black holes has a mass density $\sim10$ times greater. We therefore conclude that the observed depletion of the brightest giants in the Galactic Centre cannot be due purely to collisions of giants with MS stars or black holes. The depletion of the brightest giants may instead be due to tidal stripping of stars on very eccentric orbits, and it is also possible that modification of the stellar population by collisions of MS stars with each other may decrease the numbers of low-- and intermediate--mass giants that contribute to the bright K--band. We will look at these scenarios in detail in subsequent papers.\\
\indent We also checked to see what effect collisions would have on giants in the undepleted faint band ($15>$K$>12$). This band is dominated by $1-2$M$_{\odot}$ objects. Preventing giants of these masses evolving into this brightness band is extremely difficult, requiring the loss of $\gtrsim99\%$ of the giant's envelopes -- even with the enhanced black hole population of $2\times10^{4}$ within $0.1$\,pc, the radius to which faint band objects would be depleted is $\lesssim0.01$\,pc. We conclude that collisions with black holes and MS stars are unlikely to have an observable effect on the fainter population of giants.\\
\section{Conclusions}
Our main conclusions may be summarised as follows:\\
\indent (i) The stars missing from observations in the bands $K<10.5$ and $10.5<K<12$ in the Galactic Centre are low--mass giant stars, in the mass range $1-4$\,M$_{\odot}$.\\
\indent (ii) The Galactic Centre is a highly collisional environment. In particular, within a Galactocentric radius of $0.1$\,pc, giant stars are likely to suffer multiple collisions with both main--sequence stars and black holes.\\
\indent (iii) We have shown that significantly altering the evolution of giant stars by means of collisions requires the ejection of large quantities of mass ($>20\%$ of the envelope) while the giant is on the first giant branch. Removal of $\sim20\%$ of the envelope of a $1M_{\odot}$ giant is sufficient to make the AGB significantly fainter, but has little effect on the RGB phase. Mass losses of $\gtrsim40\%$ occurring on the RGB are sufficient to decrease the brightness of the RGB tip by $\sim1$ magnitude and prevent the star evolving onto the HB or AGB. Mass loss on the horizontal branch and AGB has essentially no effect on a star's subsequent evolution, as we showed in Figure \ref{fig:HB_AGB_mass loss}.\\
\indent (iv) We have shown that penetrating encounters with $10$\,M$_{\odot}$ black holes expel the cores of giants. The cores carry away some fraction of the envelope, resulting in very large fractional mass losses. Such mass losses in single collisions can affect the evolution of the giants to such a degree that they never evolve to become brighter than $K=12$.\\
\indent (v) We have also found, using a restricted three--body analysis, that encounters at very small $R_{\rm min}$ with $1$\,M$_{\odot}$ main--sequence impactors can also eject the giant core, leading to the severe mass losses required to prevent the giant evolving to become visible in the depleted K--bands.\\
\indent (vi) The cumulative effects of collisions with black hole and main--sequence impactors are able (for a population of black holes of $2\times10^{4}$ within $0.1$\,pc) to deplete a region $\approx 0.04$\,pc ($\approx1$\,arcsec) in radius of giants in the middle K--band, comparable in size to the region observed to be devoid of these objects by \cite{1996ApJ...472..153G}. Collisions can plausibly explain the depletion in this band.\\
\indent (vii) The cumulative effects of collisions deplete the $2-3$\,M$_{\odot}$ giants which dominate the bright K--band to a somewhat smaller radius ($\approx0.02$\,pc). This is considerably less than the size of the region observed to be depleted of these giants by \cite{1996ApJ...472..153G}. The depletion of the very brightest giants cannot be explained by collisions without invoking a black hole population so large that it would deplete the middle band out to radius much larger than observed.\\
\indent (viii) Owing to the extreme mass losses required to prevent giants evolving into this band, collisions are not likely to have an observable effect on the faint band ($15>$K$>12$).

\section{Acknowledgements}
JED gratefully acknowledges support from a Wenner--Grenn fellowship. MBD is a Royal Swedish Academy Research Fellow supported by a grant from the Knut and Alice Wallenberg Foundation. RPC gratefully acknowledges support from a Swedish Institute scholarship. The high--resolution simulations reported in Section 6 were performed at the UK Astrophysical Fluids Facility (UKAFF) at the University of Leicester, UK. All other hydrodynamical simulations were performed on machines funded by the Royal Physiographic Society, Lund. 

\bibliography{myrefs}

\label{lastpage}

\end{document}